\documentclass[12pt]{article}
\usepackage{graphicx}
\title{Fission induced by nucleons at intermediate energies}
\author{S. Lo Meo $(^{1,4})$, D. Mancusi $(^2)$, \\
C. Massimi $(^{3,4})$, G. Vannini $(^{3,4})$ and A. Ventura $(^{4})$ \\
(1) ENEA, Centro Ricerche Ezio Clementel, 40129 Bologna, Italy \\
(2) CEA, Centre de Saclay, Irfu/SPhN, F91191 Gif-sur-Yvette CEDEX, France \\
(3) Dipartimento di Fisica ed Astronomia dell' Universit\`a di Bologna, Italy \\
(4) INFN, Sezione di Bologna, 40127 Bologna, Italy }
\begin{document}
\maketitle
\begin{abstract}
Monte Carlo calculations of fission of actinides and pre-actinides induced by protons and neutrons in the energy range from 100 MeV to 1 GeV are carried out by means of a recent version of the Li\`ege Intranuclear Cascade Model, INCL++, coupled with two different evaporation-fission codes, GEMINI++ and ABLA07. In order to reproduce experimental fission cross sections, model parameters are usually adjusted on available $(p,f)$ cross sections and used to predict $(n,f)$ cross sections for the same isotopes. 
\end{abstract}

\section{Introduction}
Fission induced by nucleons at intermediate energies, \textit{i. e.} from pion production threshold ( $\sim$ 150 MeV) to a few GeV, is important from both basic and applied viewpoints. Even if fission is explained as a decay process of residual nuclei formed at the end of the fast nucleon cascade in the spallation reaction, many details are not yet clarified and deserve further experimental and theoretical work.

Important applications of intermediate energy fission are energy production with accelerator driven systems  \cite{Ma08},
radioactive waste transmutation\cite{Co10} and radiation shield design for accelerators: these applications require proton and neutron fission cross sections to be determined with high accuracy in a wide energy range.

Many experimental data have accumulated in the last sixty years: Ref.\cite{Ob01} gives a detailed review of the $(p,f)$ and $(n,f)$ measurements up to the beginning of the present century and Ref.\cite{Pr01} proposes a parametrization of $(p,f)$ cross sections based on the same experimental systematics.

Among the recent $(p,f)$ experiments a prominent role is played by Kotov et al.\cite{Ko06}, who give the cross sections tabulated in steps of 100 MeV in the range from 200 MeV to 1 GeV for several actinides very important for applications, $^{232}Th$, $^{233,235,238}U$, $^{237}Np$ and $^{239}Pu$, and for two pre-actinides, $^{nat}Pb$ and $^{209}Bi$, forming the eutectic system acting as a spallation target and as a coolant in an accelerator driven system. 

A similar role for $(n,f)$ experiments is currently being played by the n\_TOF facility\cite{Gu13} at CERN, which can measure fission cross sections from thermal energies up to about 1 GeV, with a pulsed neutron beam produced by 20 GeV/c protons from the PS accelerator impinging on a lead spallation target.
Since measuring an absolute cross section requires simultaneous determination of fission events and
neutron flux, which is a very difficult task, the measurements performed up to the present time are relative to $^{235}U(n,f)$  and absolute cross sections have been obtained by normalizing the experimental ratios to an evaluated $^{235}U(n,f)$ cross section, commonly taken from the ENDF/B-VII.1 library\cite{Ch11} up to $E_n$ = 30 MeV and from the JENDL/HE-2007 library\cite{Wa11} from 30 MeV to 1 GeV. $(n,f)$ cross sections up to 1 GeV have already been published for $^{234}U$ and $^{237}Np$\cite{Pa10}, as well as for $^{nat}Pb$ and $^{209}Bi$\cite{Ta11}. Preliminary data have been obtained for $^{232}Th$ and $^{233,238}U$.

It is experimentally known\cite{Ob01} that at the lower extremum of the intermediate energy range ( $\sim$ 100-150 MeV) the $(n,f)$ cross section is systematically lower than the $(p,f)$ cross section for a given target nucleus, the effect being larger for pre-actinides than for actinides, but the difference tends to decrease with increasing incident energy, so that at 1 GeV the behaviour of protons and neutrons is expected to be quite similar and the corresponding fission cross sections of the same order.

Main purpose of the present work is to check whether available $(p,f)$ and $(n,f)$ data for a given target nucleus can be reproduced with satisfactory accuracy using the same set of model parameters, or, at least, with very close values; in the affirmative, where only $(p,f)$ data exist in the energy range
of interest, it is reasonable to use them to predict $(n,f)$ data for the same target nucleus, or viceversa.

Our work is similar in spirit to Ref.\cite{Ba03}, where use was made of the Los Alamos codes CEM2k+GEM2
(cascade-exciton model plus generalized evaporation model) and LAQGSM+GEM2 (quark-gluon string model plus generalized evaporation model) in order to reproduce $(p,f)$ cross sections for pre-actinides and actinides, from $^{165}Ho$ to $^{239}Pu$,  taken mainly from 
Prokofiev's systematics\cite{Pr01}, extending the calculations to a large energy range, from tens of MeV to 5 GeV. 
With CEM2k+GEM2, good fits were obtained by adjusting only two parameters, the ratio, $a_f/a_n$, of level density
parameters in the fission and evaporation channels for fissioning nuclei with atomic number $70 \leq Z_f
\leq 88 $ and a constant, $C(Z_f)$, in a semi-empirical parametrization of the ratio, $ \Gamma_n/\Gamma_f$, of neutron and fission width for $Z_f \geq 89$. Using the same parameters, $(n,f)$, $(\pi,f)$ and 
$(\gamma,f)$ cross sections were calculated for the same nuclei and reasonable agreement with experimental data was obtained in several cases. With LAQGSM+GEM2, the set of parameters fitted from$(p,f)$ reactions made it possible to reproduce fission fragment production and spallation, 
fragmentation and evaporation products in heavy-ion induced reactions measured at GSI-Darmstadt in inverse kinematics\cite{Ms04}. Finally, the latest versions of the above mentioned codes, introduced as event generators in Los Alamos transport codes, such as MCNP6, where carefully tested not only 
 in the calculations of fission-fragment yields and of particles emitted from them, as is usually done with transport codes, but also in the calculations of fission cross sections\cite{Ms14}.

\section{The Models}

With the models used in the present work, fission induced by nucleons in the energy range of interest can be seen as a two-stage process:
a fast cascade stage, initiated by the high energy projectile, and representable as a succession 
of two-body collisions, with emission of fast nucleons, light clusters, pions, etc,
leaving an excited remnant, and the slow decay stage, where the remnant decays
by evaporation, fission, or other mechanisms. In our system of codes, the intranuclear cascade is described
by a recent C++ version of the Li\`{e}ge Intranuclear Cascade Model, INCL++\cite{Bo13},
the evaporation-fission model by a C++ version of GEMINI, GEMINI++\cite{Ch08}\cite{Ma10}, or a Fortran version 
of ABLA07\cite{Ke08}.

It is to be pointed out, however, that the description of intermediate energy fission as a two-
stage process is not a general rule in commonly used models and codes: many of them treat fission as a three-stage process, with an intermediate pre-equilibrium stage between fast cascade and evaporation-fission. Such an intermediate stage was originally suggested at JINR Dubna and resulted in the original version of the cascade-exciton model\cite{Gu83}, already mentioned in the introduction, which finally  evolved in the version\cite{LMa08} currently used in the Los Alamos transport codes MCNPX, MCNP6 and MARS15. The three-stage description is present in the event generators of many other well-known transport codes of general use.

Coming back to the codes used in the present work,
INCL++\cite{Bo13} is a time-like intranuclear cascade model. At the beginning of the cascade stage, the
incident nucleon is located with its own impact parameter on the surface of a working sphere, centered
on the target nucleus with a radius $R_{max} = R_0 + 8a$, where $R_0$ and $a$ are the radius and the diffuseness
of the target nucleus, respectively. Particles move along straight-line trajectories between collisions in the
working sphere and are divided into participants and spectators in the usual sense. Participants that leave the 
working sphere are considered as ejectiles. Inside the working sphere, nucleons feel a potential
that depends on energy and isospin. The depth of the potential well decreases linearly with
increasing energy, from ordinary values at the Fermi level to zero at about 200 MeV. The
isospin dependence is such that neutron and proton Fermi levels have the same energy.

Collisions are, of course, governed by Pauli blocking, treated in a different way in the first
and in the subsequent collisions. The nucleons involved in the first collision are subject to a strict
blocking: after the collision, both of them should lie outside the Fermi sphere. In subsequent collisions, 
the blocking is applied stochastically, with a probability given by the product of final-state
blocking factors. A careful definition of the latter allows one to account for surface effects
and for the depletion of the Fermi sphere during the evolution of the cascade.

An important novelty of recent versions of the code is the introduction of a coalescence model
based on phase space, which permits the emission of light clusters,
with mass $A \leq 8$, during the cascade stage, in keeping with experimental evidence.

Pions are produced in inelastic nucleon-nucleon collisions through the excitation and subsequent
decay of $\Delta$ resonances, which sets an upper limit of the order of 3 GeV to the incident
nucleon energy for the mechanism of pion production to be valid. The lower energy limit is given
by the requirement that the de Broglie wave length of relative motion be much smaller than the
range of nuclear forces, which in turn is smaller than the average distance of neighbouring 
nucleons and is commonly set to 200 MeV, although the model performs reasonably well even
at lower energies, as shown in fission calculations of the following section.

An important characteristic of the model is the self-consistent determination of the stopping time
of the cascade, which can be simply parametrized as $t_{stop} = 29.8 A_T^{0.16}$ fm/c, with $A_T$
the mass of the target nucleus. At $t = t_{stop}$ many physical quantities, such as the excitation
energy of the target nucleus and the average kinetic energy of the ejectiles, switch from a fast time evolution, dominated by intranuclear cascade, to a much slower evolution, which is taken as a signature of equilibration. Thanks to this choice of the stopping
time, it is not necessary to introduce a pre-equilibrium model describing the intermediate stage
between the fast cascade and the evaporation-fission decay. The effect of an explicit pre-equilibrium stage in the INCL model on nucleon-induced reactions above 200 MeV was tested in Ref.\cite{Cu97} and found not really necessary. The effect might be more significant in the energy range from 100
to 200 MeV, with particular reference to energy spectra and angular distributions of emitted particles, rather than the fission cross sections considered in the present work. However, no clear evidence in this
sense emerged from the latest IAEA Benchmark of Spallation Models\cite{Sa10}, whose conclusions are presented in Ref.\cite{Le11}.

GEMINI++\cite{Ch08}\cite{Ma10} is a statistical-model code which follows the decay of a compound nucleus by a series of
sequential binary decays until such decays are forbidden by energy conservation or become improbable because
of gamma-ray competition. Differently from most statistical-model codes, light-particle evaporation
is described by the Hauser-Feshbach formalism, which strictly conserves angular momentum, at the
price of higher computational time with respect to the more common Weisskopf-Ewing formalism.
An important ingredient of the decay width is the nuclear level density as a function of excitation
energy $U$ and angular momentum $J$, described by a Bethe-type formula
\begin{equation}
\rho(E^*,J) \sim (2J+1) \exp\left[2 \sqrt{a\left(U\left(E^*\right),J\right)U} \right] \quad.
\label{Bethe}
\end{equation}
Here, $E^*$ is the excitation energy and the thermal excitation energy $U$ is related to $E^*$ by the formula
\begin{equation}
U= E^* - E_{yrast}(J) - \delta P - \delta W \quad ,
\label{excit}
\end{equation}
where $E_{yrast}(J)$ is the yrast energy for angular momentum $J$, $\delta P$ and $\delta W$ are the pairing and shell correction, respectively, which depend on excitation energy.
The level density parameter $a(U,J)$ includes the damping of shell correction $\delta W$ inspired
by Ignatyuk\cite{Ig75}
\begin{equation}
a(U)=\tilde a (U) \left[ 1- h\left(\frac{U}{\eta}+\frac{J}{J_\eta}\right)\frac{\delta W}{U}\right] \quad ,
\label{a_Ig}
\end{equation}
with $h(x)=\tanh (x)$, $\eta$ = 19 MeV and $J_\eta$ = 50. The effective level density parameter $\tilde a (U)$ is taken of the form\cite{Ch10}
\begin{equation}
\tilde a \left( U \right) = \frac{A}{k_\infty -\left(k_\infty-k_0\right)\exp\left(-\frac {\gamma}{k_\infty-k_0}\frac{U}{A}\right)}\quad ,
\label{a1}
\end{equation}
with $k_0$ = 7.3 MeV, $k_\infty$ = 12 MeV and $\gamma = 0.00517\exp\left(0.0345 A\right)$, with $A$ the mass number. 

The decay width for symmetric fission, dominant at high excitation energy, is given by the well-known Bohr-Wheeler formula, while
the decay width for asymmetric fission is derived from Moretto's formalism\cite{Mo75} \cite{Mo75bis}, based on the concept of conditional fission barrier, \textit{i. e.}
a saddle point configuration with fixed asymmetry of mass and charge of the prefragments.
Liquid drop barriers are calculated by means of Sierk's finite-range model\cite{Si86}, with shell and pairing corrections
taken from Ref.\cite{Mo95}.

An important
adjustable parameter in the code is the ratio of the effective level density parameter at the saddle point, $\tilde a_f$, to the same
quantity at ground-state deformation, $\tilde a_n$, with a default value of 1.036\cite{Ma10}. Fission transients are not considered, although it has been shown\cite{Ma10} that the model can accommodate a short fission delay.

ABLA07\cite{Ke08} is a statistical code describing the de-excitation of a nucleus in thermal equilibrium by particle evaporation, fission, or, above
a prescribed excitation energy per nucleon, multifragmentation. Particle evaporation is treated in an extended Weisskopf-Ewing formalism, where
a distribution of orbital angular momenta in the emission of nucleons and clusters is evaluated in semiclassical approximation, based on phase
space arguments. The model also allows generalized evaporation of excited clusters, which plays a role analogous to asymmetric fission in GEMINI++.

An essential ingredient of the decay formalism is the nuclear level density as a function of excitation energy and
angular momentum, described by a constant temperature formula at low energy and by a Bethe-type formula (\ref{Bethe}) at high energy.
The asymptotic level density parameter, $\tilde a$ (see formula (\ref{a_Ig})), is energy-independent and given, in MeV$^{-1}$, by the original prescription
of Ref.\cite{Ig75}
\begin{equation}
\tilde a = 0.073 A + 0.095 B_s A^{2/3} \quad ,
\label{a2}
\end{equation}
where $B_s$ is the ratio of the surface of the deformed nucleus to that of a spherical nucleus with the same mass number $A$, as in the finite-range liquid drop model\cite{Si86}. The level density
contains a collective enhancement factor, due to nuclear rotations and vibrations, depending on excitation energy\cite{Ju98}.
The approach to fission contains elements of dynamics: the time evolution of the fission degree of freedom is treated as a diffusion process, determined by the
interaction of collective degrees of freedom with a heat bath formed by the individual nucleons and described by a Fokker-Planck equation whose
solution leads to a time-dependent fission width, $\Gamma_f(t)$. An analytical approximation to such a solution and, consequently, to the time
dependence of the fission width makes the problem computationally tractable.
At low excitation energy, the code uses the standard model of a two-humped fission barrier, whose penetrability is computed in the approximation
of full damping of the vibrational resonances in the intermediate well. Like in the GEMINI++ code, liquid drop barriers are computed in the
frame of the finite-range model\cite{Si86} and shell and pairing corrections are taken from Ref.
\cite{Mo95}.

\section{Fission Cross Sections}
As already stressed in the introduction, our main purpose is a simultaneous reproduction of $(p,f)$ and $(n,f)$ cross sections for the same target nuclei, possibly using the same, or, at least, rather close values of model parameters. In fact, it is expected that the isospin dependence of the fission cross section is already contained in the models we use.

In order to better understand how fission keeps memory of projectile isospin, it is worth recalling that the fission cross section increases with increasing fissility parameter, $x \simeq \frac{1}{49}\frac{Z^2}{A}$, where $Z$ and $A$ are charge and mass of the fissioning nucleus. In the case of a proton induced reaction, the highest fissility parameter is obtained when the incident proton is captured by the target nucleus and is proportional to ${(Z_T+1)^2}/{(A_T+1)}$, with $Z_T$ and $A_T$ the charge and mass of the target nucleus, while in the case of capture of an incident neutron the fissility parameter is smaller, since it is proportional to ${Z_T^2}/{(A_T+1)}$.

An analysis of charge and mass distributions of remnants undergoing fission at the end of the fast cascade stage, carried out in Ref.\cite{Bu00} with the cascade-exciton model code CEM95\cite{Ma95} for $(p,f)$ and $(n,f)$ reactions up to a projectile energy $E_{proj} $ = 200 MeV showed the relevance of remnants with charge $Z_T+1$ in enhancing the $(p,f)$ cross section with respect to the $(n,f)$ cross section, where remnants cannot have charges larger than $Z_T$. By considering the fissility parameter dependence of the liquid drop barriers, the authors of Ref.\cite{Bu00} explained also why the difference of $(p,f)$ and $(n,f)$ cross sections is much larger in pre-actinides than in actinides. 
It is worth recalling that a CEM95 analysis and interpretation of the differences between $(p,f)$ and $(n,f)$ cross sections
for $^{208}Pb$ and $^{209}Bi$ in the energy range from 45 to 500 MeV had already been published in
Ref.\cite{Pr99}.

The analysis of differences of $(p,f)$ and $(n,f)$ reactions was further enriched with new experimental data in the pre-actinide region by the authors of Ref.\cite{Sm01}, who derived also an approximate analytical dependence of the ${\sigma_{pf}}/{\sigma_{nf}}$ ratio on ${Z_T^2}/{A_T}$ at given projectile energy, $E_{proj}$. For instance, at $E_{proj} \simeq$ 180 MeV, they obtained 
\begin{equation}
\frac{\sigma_{pf}}{\sigma_{nf}}\simeq \exp\left[ 0.26\left(36.6-\frac{Z_T^2}{A_T}\right)\right]\quad ,
\label{sigma_ratio}
\end{equation}
valid for both actinides and pre-actinides.
With increasing projectile energy up to the GeV region, it is expected that the differences in the behaviour of neutrons and protons become smaller, so that $(p,f)$ and $(n,f)$ cross sections tend to a common value.   

In keeping with the philosophy of the authors of the INCL++ code\cite{Bo13}\cite{Ma14}, no parameters have been modified in the intranuclear cascade model, already optimized by reproducing a large amount of data
related to the cascade stage, such as total reaction cross sections, double-differential spectra of emitted nucleons, pions and light clusters, while two basic fission model parameters have been taken as adjustable either in GEMINI++ or ABLA07: the height of the liquid-drop fission barrier, $B_f$, and the asymptotic level density parameter at the saddle point, $\tilde a_f$, in ABLA07, or the ratio of the level density parameter at the saddle point to that at the ground state deformation, $\tilde a_f/\tilde a_n$, in GEMINI++. Both parameters affect significantly the calculation of fission cross sections in the whole energy range of interest to the present work, from 100 MeV to 1 GeV, although the change of the barrier height is more important at low incident energy and the change of the level density parameter at high energy. In principle, many sets of experimental data could be reproduced by modifying only one parameter, \textit{e.g.} $\tilde a_f$, at the the cost of using rather different values of it in the neutron and proton channels, particularly for actinides, as we did in our preliminary work\cite{Lo13}.

In the following sub-section results for absolute $(p,f)$ and $(n,f)$ cross sections are presented and discussed in comparison with experimental data, while a separate sub-section will be dedicated to $(n,f)$ cross sections relative to $^{235}U(n,f)$, with particular reference to the comparison with n\_TOF data.

\subsection{Absolute Cross Sections}

The rationale behind our work is to adjust fission model parameters on the $(p,f)$ cross sections of Ref.\cite{Ko06} for $^{nat}Pb$, $^{209}Bi$, $^{232}Th$, $^{233,235,238}U$ and $^{239}Pu$ and use similar parameter values in computing $(n,f)$ cross sections for the same targets.
As alredy pointed out in our preliminary work\cite{Lo13}, the GEMINI++ and ABLA07 models can produce fits of comparable accuracy for actinides, while in the lead-bismuth region ABLA07 appears to work better than GEMINI++. Therefore, we have decided to show the results of GEMINI++ calculations for actinides and those of ABLA07 calculations for lead and bismuth.

The experiment of Ref.\cite{Ko06}, labelled as Kotov 2006 in the figures, yields 9 cross section values in the energy range from 207 MeV to 1 GeV. We have evaluated the quality of our fits by means of a $\chi^2$ test, where the $\chi^2$ variable is defined as
\begin{equation}
\chi^2_0 = \sum_{i=1}^9\left(\frac{{\sigma_i}_{calc.}-{\sigma_i}_{exp.}}{\Delta{\sigma_i}_{exp.}}\right)^2\quad,
\label{chi_sq}
\end{equation}
where the symbols are self-explanatory. Since we try to reproduce 9 experimental values, ${\sigma_i}_{exp.}$, by adjusting 2 parameters, the relevant $\chi^2$ distribution, $f_\nu\left(\chi^2\right)$, should have $\nu$ = 9-2 degrees of freedom and the quality of fit is evaluated by means of the cumulative probability
\begin{equation}
Q\left(\chi^2_0\right) = \int_{\chi^2_0}^\infty f_7\left(\chi^2\right)d\chi^2 \quad .
\label{cum_prob}
\end{equation}

\subsubsection{Actinides}

Table \ref{tabI} shows the values of the parameters adjusted on the experimental $(p,f)$ data of Ref.\cite{Ko06} for actinides in our INCL++/GEMINI++ calculations, \textit{i. e.} the ratio of the level density parameter at the saddle point to the one at ground-state deformation, $\tilde a_f/\tilde a_n$, and the global correction of the liquid-drop fission barriers, $\Delta B_f$, in MeV. 

\begin{table}[h]
	\begin{center}
		\begin{tabular}{|c|c|c|c|c|c|}\hline
			Isotope    & $Z^2/A$ & $\tilde a_f/\tilde a_n$ & $\Delta B_f$ (MeV) & $\chi^2_0$ & $Q\left(\chi^2_0\right)$ \\ \hline
			$^{232}Th$ & 34.91   &  1.040    &  -0.3              &  3.71      &    0.81           \\ \hline
			$^{238}U$  & 35.56   &  1.038    &  -0.3              &  4.50      &    0.72            \\ \hline
			$^{235}U$  & 36.02   &  1.050    &  -0.2              &  2.75      &    0.91            \\ \hline
			$^{233}U$  & 36.33   &  1.100    &  -0.5              &  4.16      &    0.76             \\ \hline
			$^{237}Np$ & 36.49   &  1.040    &   0.0              & 25.01      &    0.0007           \\ \hline
			$^{239}Pu$ & 36.97   &  1.036    &  +0.5              &  8.44      &    0.30             \\ \hline      
		\end{tabular}
	\end{center}
	\caption{Model parameters and $\chi^2$ tests for proton-induced fission of actinides.}
	\label{tabI}
\end{table}

\subsubsection{$^{235}U$}
\begin{figure}[htb]
	\centering
	\includegraphics[width=12cm]{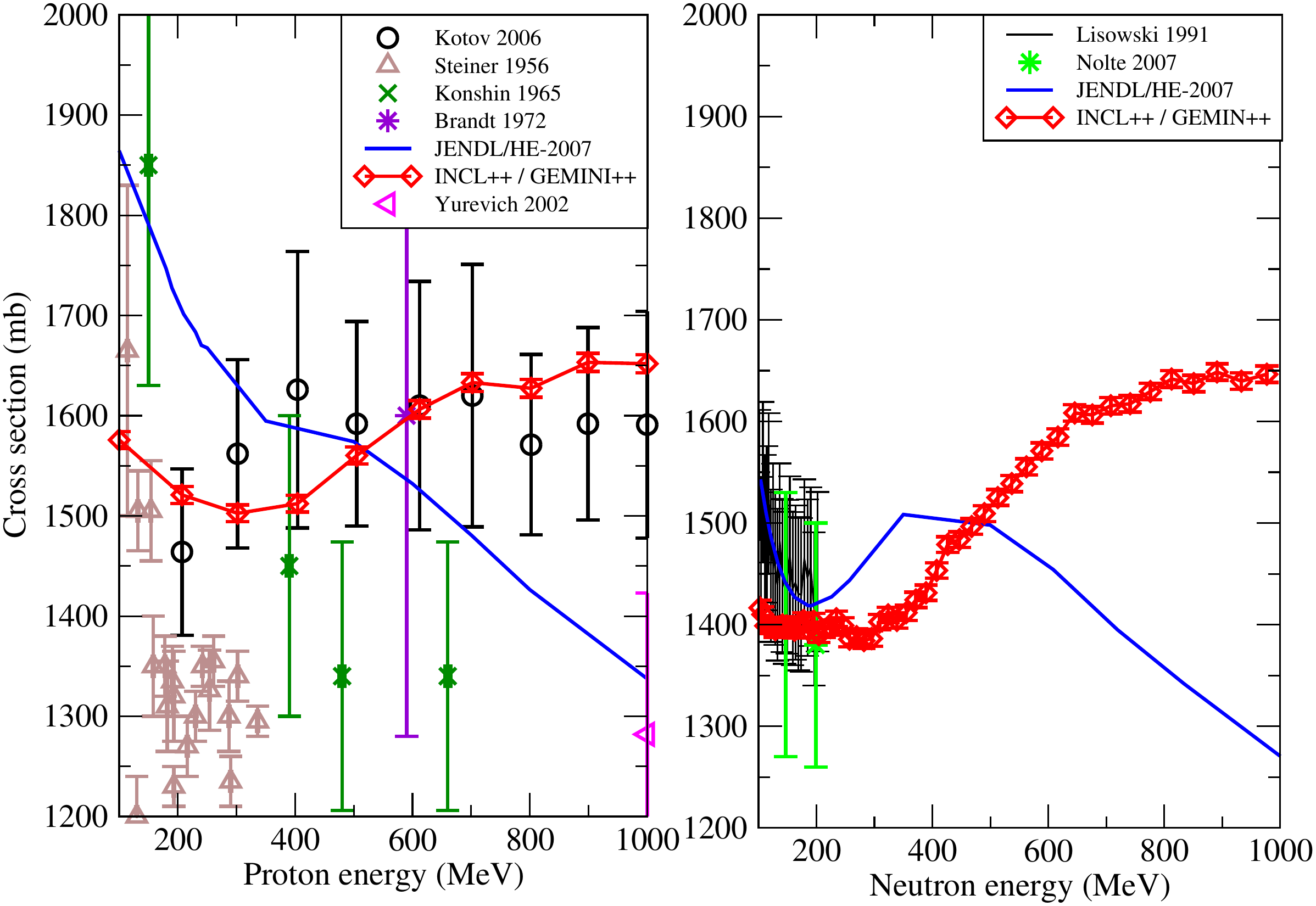}
	\caption{ $^{235}U$ fission cross sections. Left panel: $(p,f)$; right panel: $(n,f)$. Experimental data are discussed in the text.}
	\label{label1}
\end{figure}

Fig. \ref{label1} shows calculated $(p,f)$ and $(n,f)$ cross sections for $^{235}U$ in comparison with experimental data.

As far as $(p,f)$ data are concerned, our reference values\cite{Ko06} and, consequently, our calculated cross section appear to be compatible with the old measurement of Ref.\cite{St56} (Steiner 1956) only below 150 MeV and in agreement with the experimental point at 590 MeV from Ref.\cite{Br72} (Brandt 1972), but in conflict with Ref.\cite{Ko65} (Konshin 1965), which seems to have inspired the JENDL/HE-2007 evaluation, together with the more recent point at 1 GeV from Ref.\cite{Yu02} (Yurevich 2002). 

Reliable $(n,f)$ data are available only below 200 MeV and are taken from Refs.\cite{Li91} (Lisowski 1991) and \cite{No07} (Nolte 2007) , which are in mutual agreement; in both experiments fission events were detected simultaneously with $(n,p)$ scattering events, so that from knowledge of the differential $(n,p)$ scattering cross section it was possible to determine the neutron flux and, consequently, the absolute fission cross section. It is worth recalling that on the basis of the data of Ref.\cite{Li91} it was proposed in Ref.\cite{Ca97} to extend to 200 MeV the energy range where the $^{235}U(n,f)$ cross section can be considered as a fission standard.  A previous measurement up to 750 MeV\cite{Ra87} yielded cross section values much lower than those of Refs.\cite{Li91}\cite{No07} and has not been taken into account in this work.

The data of Refs.\cite{Li91},\cite{No07} are reproduced by JENDL/HE-2007 and by us.
Differently from the JENDL evaluation, however, our $(p,f)$ and $(n,f)$ cross sections show a large plateau above 500 MeV and tend to a common value at 1 GeV. It is worth pointing out that our $(p,f)$ and $(n,f)$ calculations are done with the same model parameters, namely $\tilde a_f / \tilde a_n $ = 1.050 and a common reduction of all the barrier heights of remnants in both processes by an amount $\Delta B_f $ = -0.2 MeV.


\subsubsection{$^{238}U$}

\begin{figure}[h]
	\centering
	\includegraphics[width=12cm]{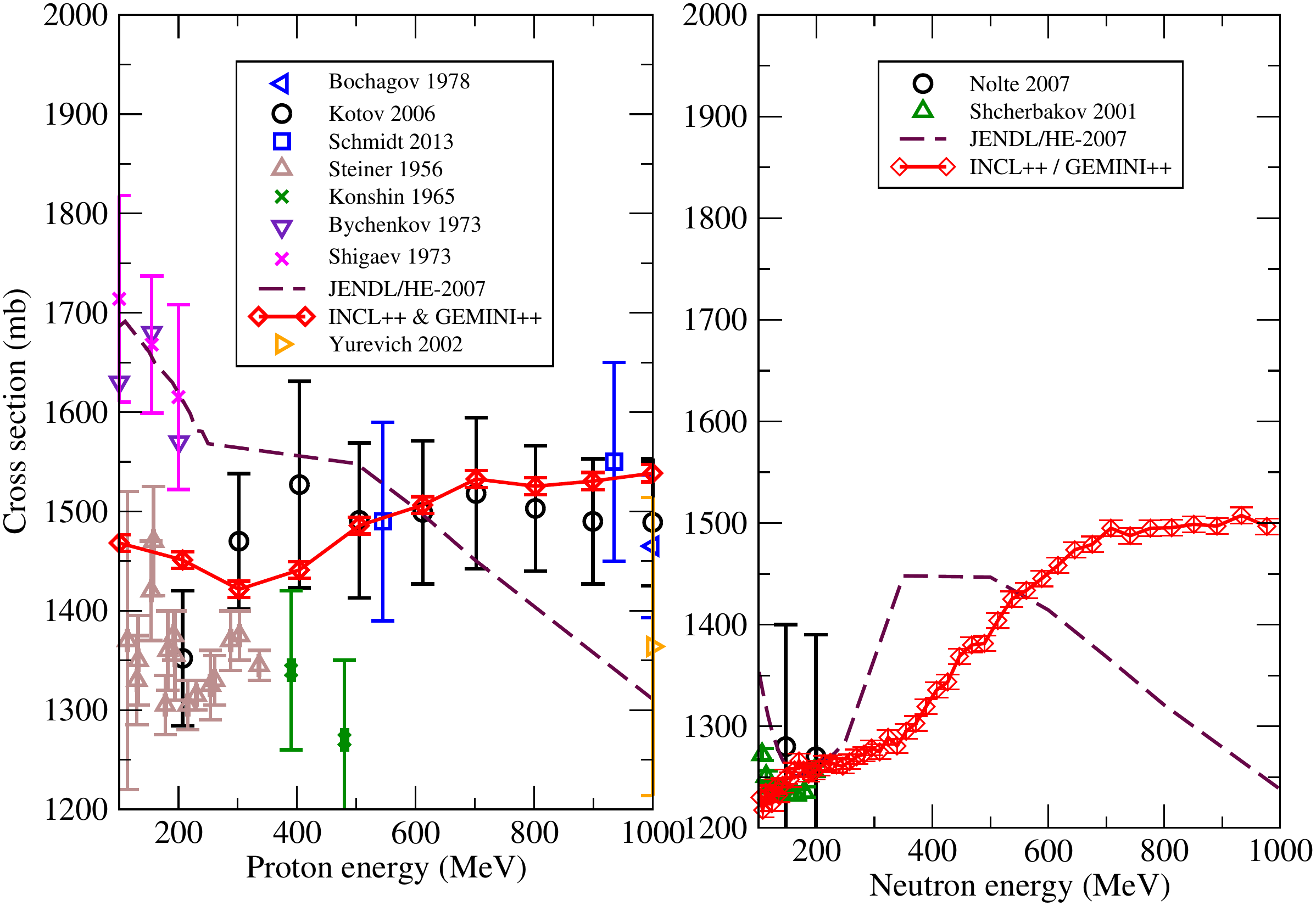}
	\caption{ $^{238}U$ fission cross sections. Experimental data are discussed in the text.}
	\label{label2}
\end{figure}

Similar calculations, done for $^{238}U$, are shown in Fig. \ref{label2}.
Our $(p,f)$ cross section, adjusted on Kotov's data\cite{Ko06}, appears to be compatible with the old data of Ref.\cite{St56} below 150 MeV and with the recent data of Ref.\cite{Sc13} (Schmidt 2013), obtained in an inverse-kinematics experiment, at 545 and 935 MeV, as well as with the older value at 1 GeV from Ref.\cite{Bo78} (Bochagov 1978). On the contrary, the JENDL/HE-2007 evaluation is consistent with Refs.\cite{Sh73},\cite{By73} (Shigaev 1973 and Bychenkov 1973), as well as with Kotov's data in the intermediate range, up to 600 MeV, and the value at 1 GeV from Ref.\cite{Yu02} (Yurevich 2002).

Experimental $(n,f)$ data are available below 200 MeV: as already mentioned in connection with $^{235}U$, the data of Ref.\cite{No07} can be considered as a true absolute cross section, while the data of Ref.\cite{Sh01} (Shcherbakov 2001) are normalized to the $^{235}U(n,f)$ cross section recommended in Ref.\cite{Ca97b}. We are in good agreement with Ref.\cite{No07}, while underestimating the data of Ref.\cite{Pr96} (Prokofiev 1996) and overestimating the old data\cite{Go55} (Goldanskii 1955). Again, our calculations are done with the same parameters for
the $(p,f)$ and $(n,f)$ reactions, namely $\tilde a_f/\tilde a_n$ = 1.038 and $\Delta B_f$ = -0.3 MeV.

\subsubsection{$^{nat}U$}

\begin{figure}[h]
	\centering
	\includegraphics[width=11cm]{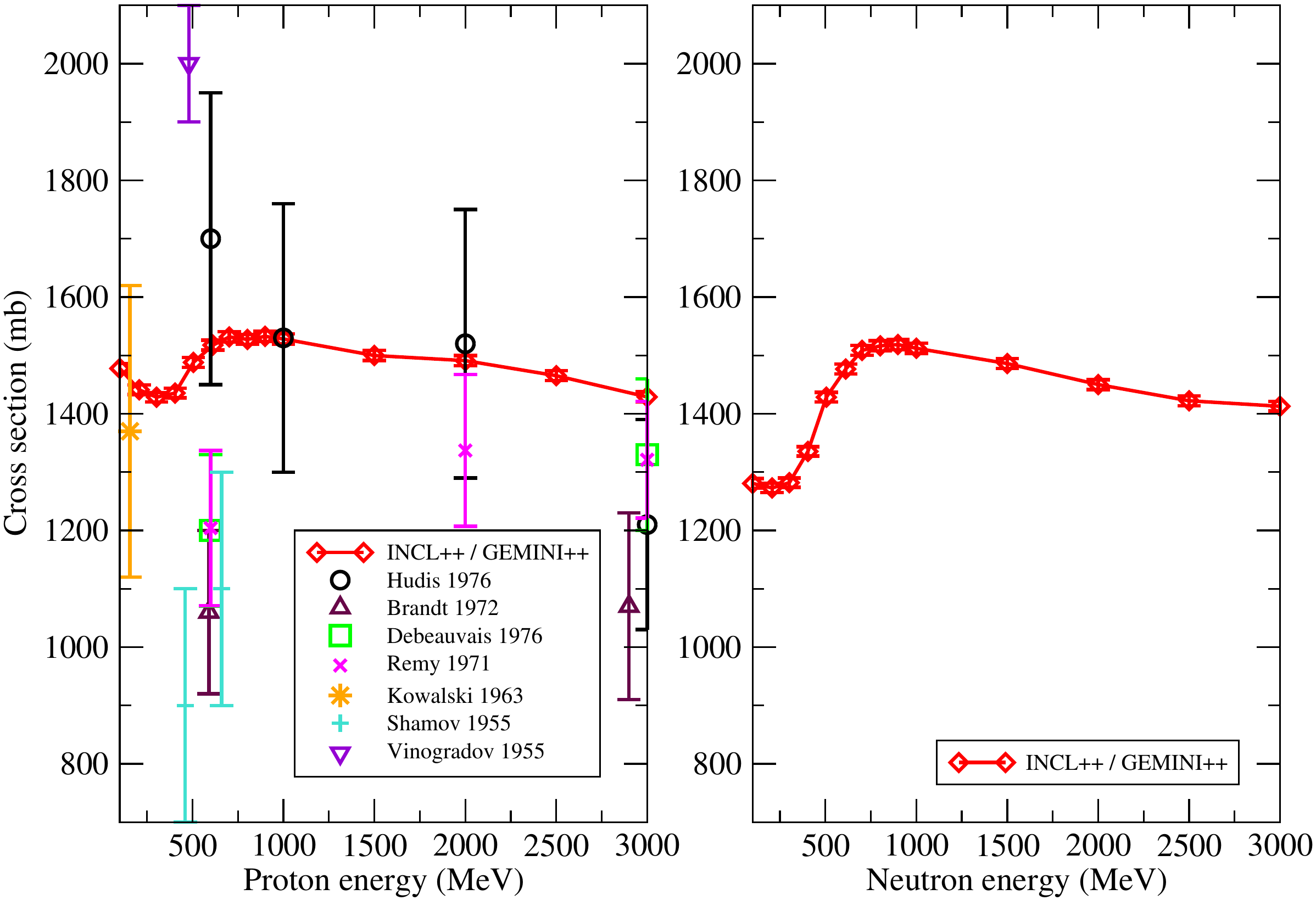}
	\caption{ $^{nat}U$ fission cross sections. Experimental data are discussed in the text.}
	\label{label2bis}
\end{figure}

As shown in Figs. (\ref{label1},\ref{label2}), the JENDL/HE-2007 evaluations of fission cross sections decrease with increasing projectile energy much faster than our calculations: a possible explanation for the discrepancy is that in the JENDL evaluations the energy dependence of fission cross sections is not derived from a full numerical simulation, but estimated by an analytical formula\cite{Fu02} with parameters based on systematics of experimental data, which are mainly available below 200 MeV. It is then of interest to check whether experimental indications of either trend exist at energies between 1 GeV and 3 GeV, which
is at the same time the upper limit of JENDL/HE-2007 evaluations and of the validity of the intranuclear cascade model in the version of INCL++ we use. This possibility is provided by natural uranium, not explicitly evaluated in the JENDL library, but expected, in this range of projectile energy, to be very similar to $^{238}U$, its main component, with an abundance of more than 99\%. The left panel of Fig. \ref{label2bis} shows the $^{nat}U(p,f)$ cross section calculated between 100 MeV and 3 GeV with the same parameters as $^{238}U$, namely $\tilde a_f / \tilde a_n$ = 1.038 and $\Delta B_f$ = -0.3 MeV.

 Our calculation compares well with the experimental data of Ref.\cite{Hu76} (Hudis 1976), which cover the full energy range, and is in marginal agreement with the point at 3 GeV, where, on the other hand, our calculation agrees with the data from Ref.\cite{DB76} (Debeauvais 1976), Ref.\cite{Re71} (Remy 1971) and Ref.\cite{Sh55} (Shamov 1955); in turn, the last three experiments are lower than Hudis 1976 and our calculations at lower energies. We are also in agreement with the point  from Ref.\cite{Ko63} (Kowalski 1963).

 Finally, all the data from Ref.\cite{Br72} (Brandt 1972) are lower than our calculations, while the old experiment of Ref.\cite{Vi55} is much larger. Summing up, we believe that our calculations of the $^{nat}U(p,f)$ reactions are supported by experimental evidence up to 3 GeV. For the sake of completeness, the right panel of Fig.\ref{label2bis} shows the $^{nat}U(n,f)$ cross section, calculated with the same model parameters.

\subsubsection{$^{233}U$}

\begin{figure}[h]
	\centering
	\includegraphics[width=12cm]{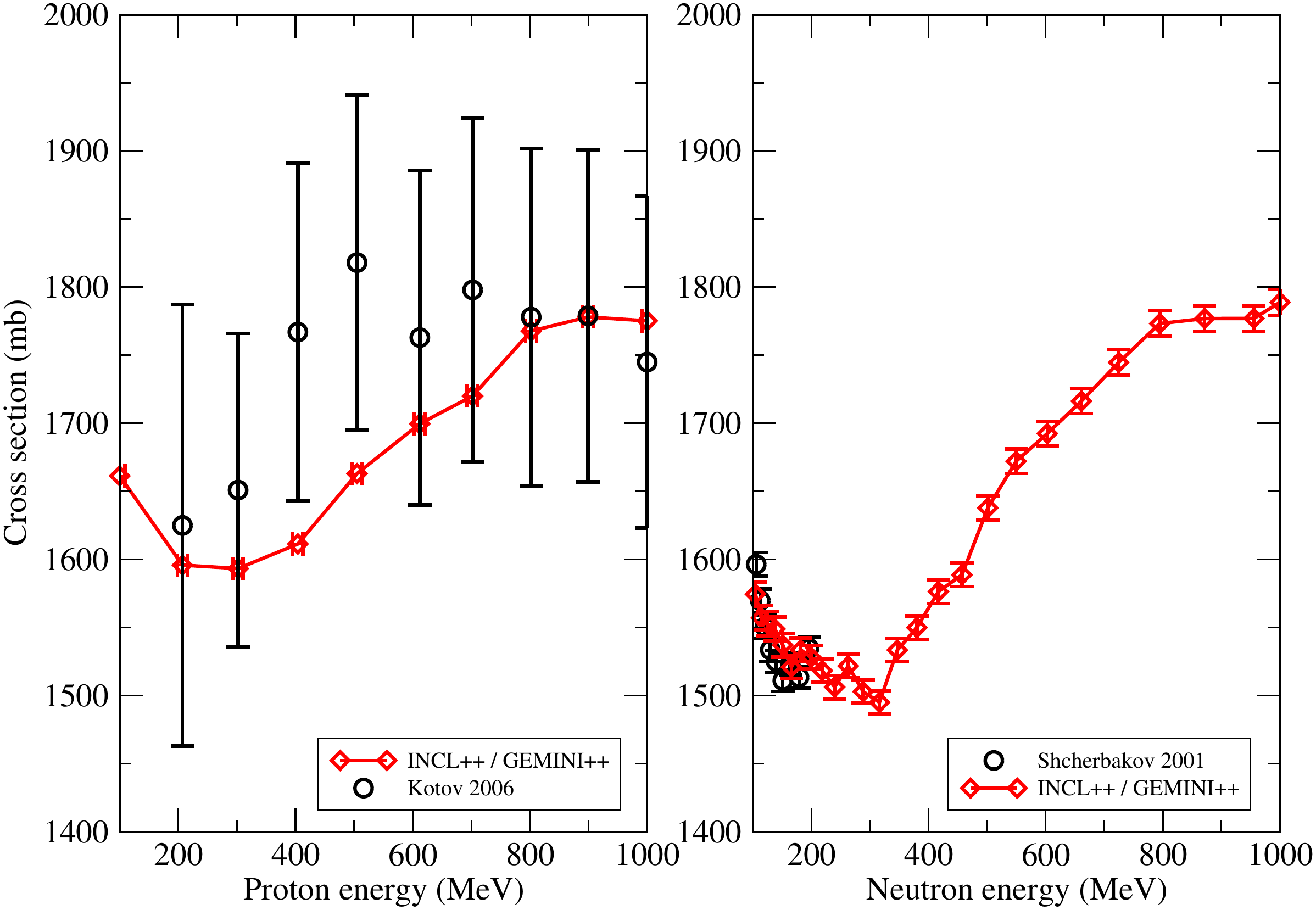}
	\caption{ $^{233}U$ fission cross sections. Experimental data are discussed in the text.}
	\label{label3}
\end{figure}

The only experimental data of the $^{233}U(p,f)$ reaction in the energy range of interest are those of Kotov\cite{Ko06}, shown in the left panel of Fig.\ref{label3} together with our theoretical fit, which underestimates the two points at 404 and 505 MeV. In order to reproduce them, one should resort to an abnormally high value of the $\tilde a_f/\tilde a_n$ ratio, such as 1.6 or more. We have preferred to assume $\tilde a_f/\tilde a_n$ = 1.10, higher than the values adopted for $^{235}U$ and $^{238}U$, but of the same order as that of the original fit of Ref.\cite{Ko06}, which, in addition, points out the difficulty in reproducing $(p,f)$ cross sections of the uranium chain with their own cascade-exciton-fission model. The price we pay for this reasonable value of $\tilde a_f/\tilde a_n$ is a large reduction of the liquid-drop fission barriers of remnants, $\Delta B_f$ = -0.5 MeV.


The $^{233}U(n,f)$ cross section relative to $^{235}U$ has been measured at n\_TOF up to $E_n\simeq$ 1 GeV, but final data are not yet available. Other relative $(n,f)$ measurements in a smaller energy range are those of Refs.\cite{Sh01},\cite{Li91b}. The right panel of Fig.\ref{label3} shows that available $(n,f)$ data can be reproduced by a model calculation with $\tilde a_f/\tilde a_n$ = 1.10, as for the $(p,f)$ reaction, but $\Delta B_f$ = 0.  A more detailed comparison of experimental and theoretical cross section ratios will be shown in the following sub-section.

\subsubsection{$^{232}Th$}

\begin{figure}[h]
	\centering
	\includegraphics[width=12cm]{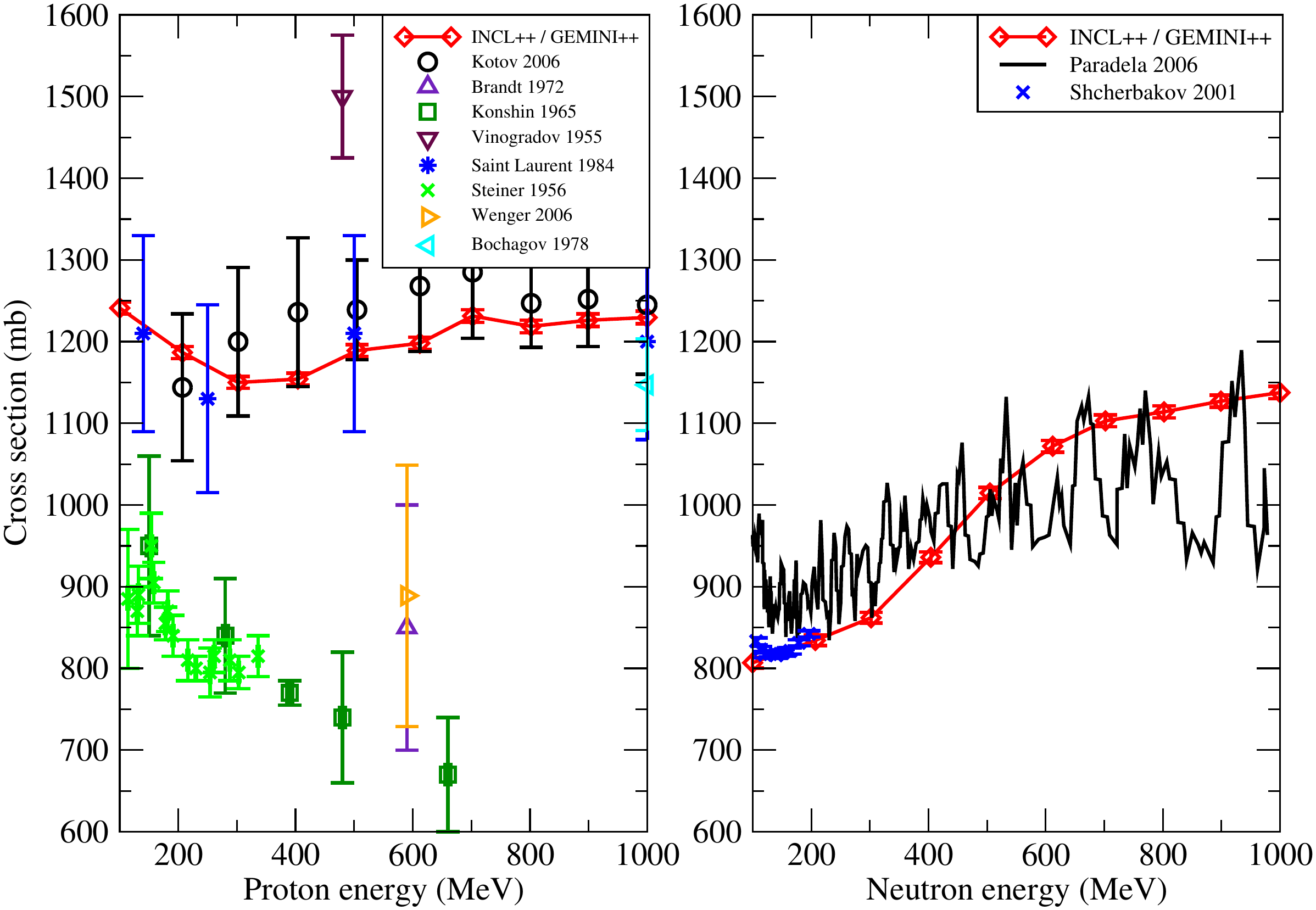}
	\caption{ $^{232}Th$ fission cross sections. Experimental data are discussed in the text.}
	\label{label4}
\end{figure}

As shown the left panel of Fig.\ref{label4}, the experimental results available for the $(p,f)$ cross section of $^{232}Th$ can be classified in two series: our reference experiment\cite{Ko06}, Ref.\cite{Bo78} (Bochagov 1978) and Ref.\cite{SL84} (Saint-Laurent 1984) on one side and, on the other side,  the majority of older experiments, with the exception 
of Ref.\cite{Vi55} (Vinogradov 1955), but including also the recent measurement\cite{We06} (Wenger 2006) at 590 MeV, which yield systematically lower values.
Our choice between the two series of $(p,f)$ measurements has been motivated also by considerations on the experimental $(n,f)$ cross section, which is expected to be significantly smaller than the $(p,f)$ cross section in the 100 MeV range: in particular, formula (\ref{sigma_ratio}), based on the systematics 
of Ref.\cite{Sm01}, predicts for $^{232}Th$ $\sigma_{pf}/\sigma_{nf}$ = 1.55 at $E_{proj} \simeq$ 180 MeV. A look at the experimental $(n,f)$ cross sections of Ref.\cite{Pa06} (Paradela 2006) and Ref.\cite{Sh01} (Shcherbakov 2001) induces us to prefer the series of higher experimental values\cite{Ko06}\cite{SL84}, otherwise the $(p,f)$ and $(n,f)$ cross sections at low energy would be more or less of the same order.

It is to be stressed, however, that both $(n,f)$ measurements\cite{Sh01}\cite{Pa06} are relative to $^{235}U$ and that absolute cross sections have been obtained with different criteria of normalization; moreover, the n\_TOF data\cite{Pa06} are still preliminary and subject to possible changes. As a consequence, the present calculation of the $(n,f)$ cross section should be taken as preliminary, too.
The $(p,f)$ cross section has been reproduced with $\tilde a_f/\tilde a_n$ = 1.040 and $\Delta B_f$ = -0.3 MeV, the
$(n,f)$ cross section with the same ratio of level density parameters, but with $\Delta B_f$ = +0.1 MeV, which is a relatively large discrepancy. If final $(n,f)$ values from the n\_TOF experiment turn out to be higher than the preliminary ones, the discrepancy will be hopefully reduced.

\subsubsection{$^{237}Np$}

\begin{figure}[h]
	\centering
	\includegraphics[width=10cm]{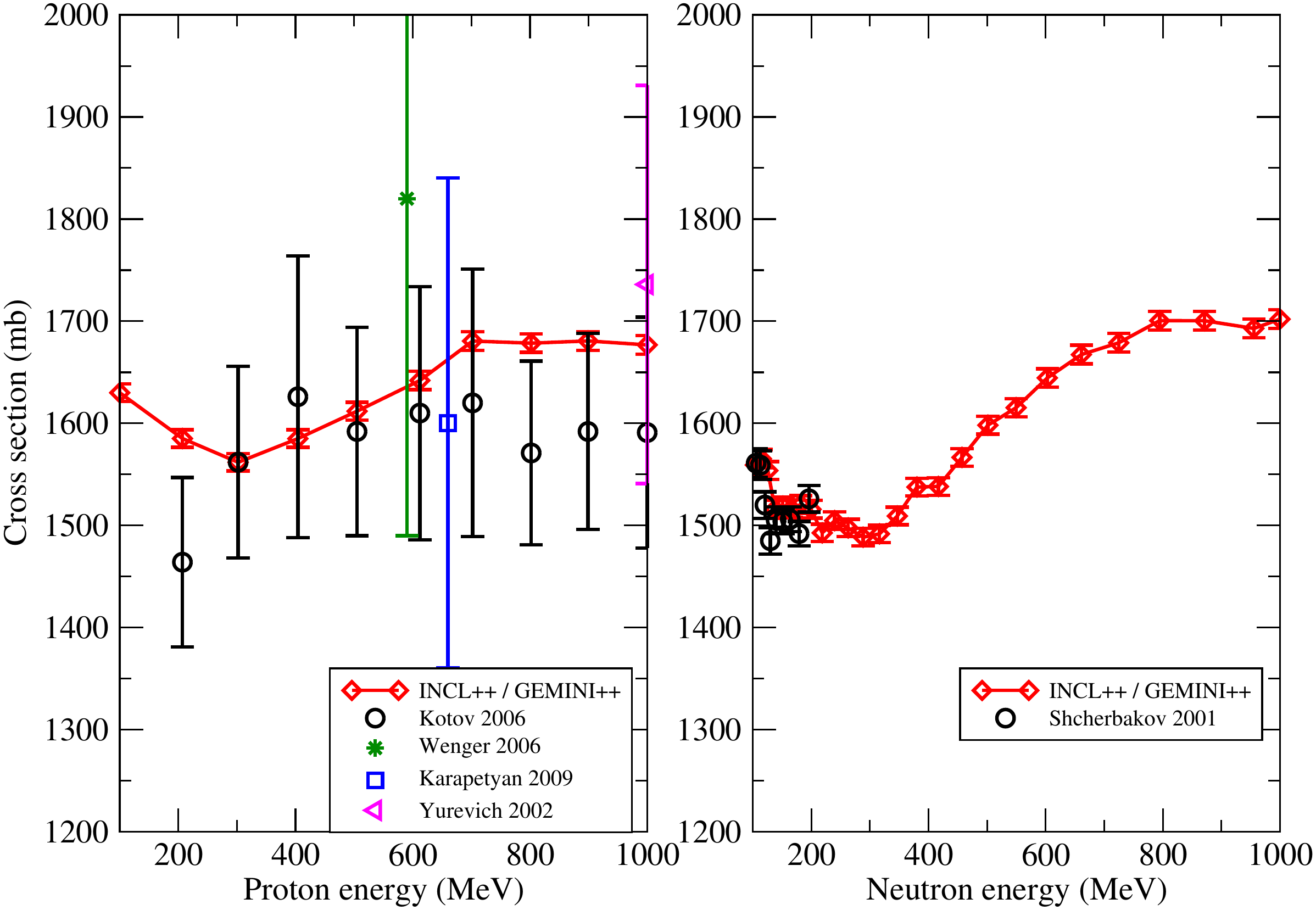}
	\caption{ $^{237}Np$ fission cross sections. Experimental data are discussed in the text.}
	\label{label5}
\end{figure}

The basic measurement of the $^{237}Np(p,f)$ cross section in the energy range of interest is again Kotov 2006\cite{Ko06}, to which one can add a point at 590 MeV from Ref.\cite{We06} (Wenger 2006), a point at 660 MeV from Ref.\cite{Ka09} (Karapetyan 2009) and a point from Ref.\cite{Yu02} (Yurevich 2002) at 1 GeV; all of them agree with Kotov's data within the experimental uncertainties. The theoretical fit shown in the left panel of Fig. \ref{label5} is characterized by a large $\chi^2_0$ value, mainly because of the point at 207 MeV, which is largely overestimated by the calculation. .
A similar problem appears in the original fit of Ref.\cite{Ko06}, which reproduces data below 100 MeV from other experiments, clearly not compatible with the low value of the $(p,f)$ cross section at 207 MeV
Using the same model parameters, $\tilde a_f/ \tilde a_n$ = 1.040 and $\Delta B_f$ = -0.2 MeV, for the $^{237}Np(n,f)$ reaction, one obtains the cross section shown in the right panel of Fig.\ref{label5}, which reproduces the data of Ref.\cite{Sh01} (Shcherbakov 2001) below 200 MeV. It is to be recalled, however, that the measurements of Ref.\cite{Sh01} are relative to $^{235}U(n,f)$ and that other measurements of relative cross sections exist, including those from n\_TOF\cite{Pa10}, which cover the full energy range of interest: they deserve a more detailed description in the following sub-section.

\subsubsection{$^{239}Pu$}

\begin{figure}[h]
	\centering
	\includegraphics[width=10cm]{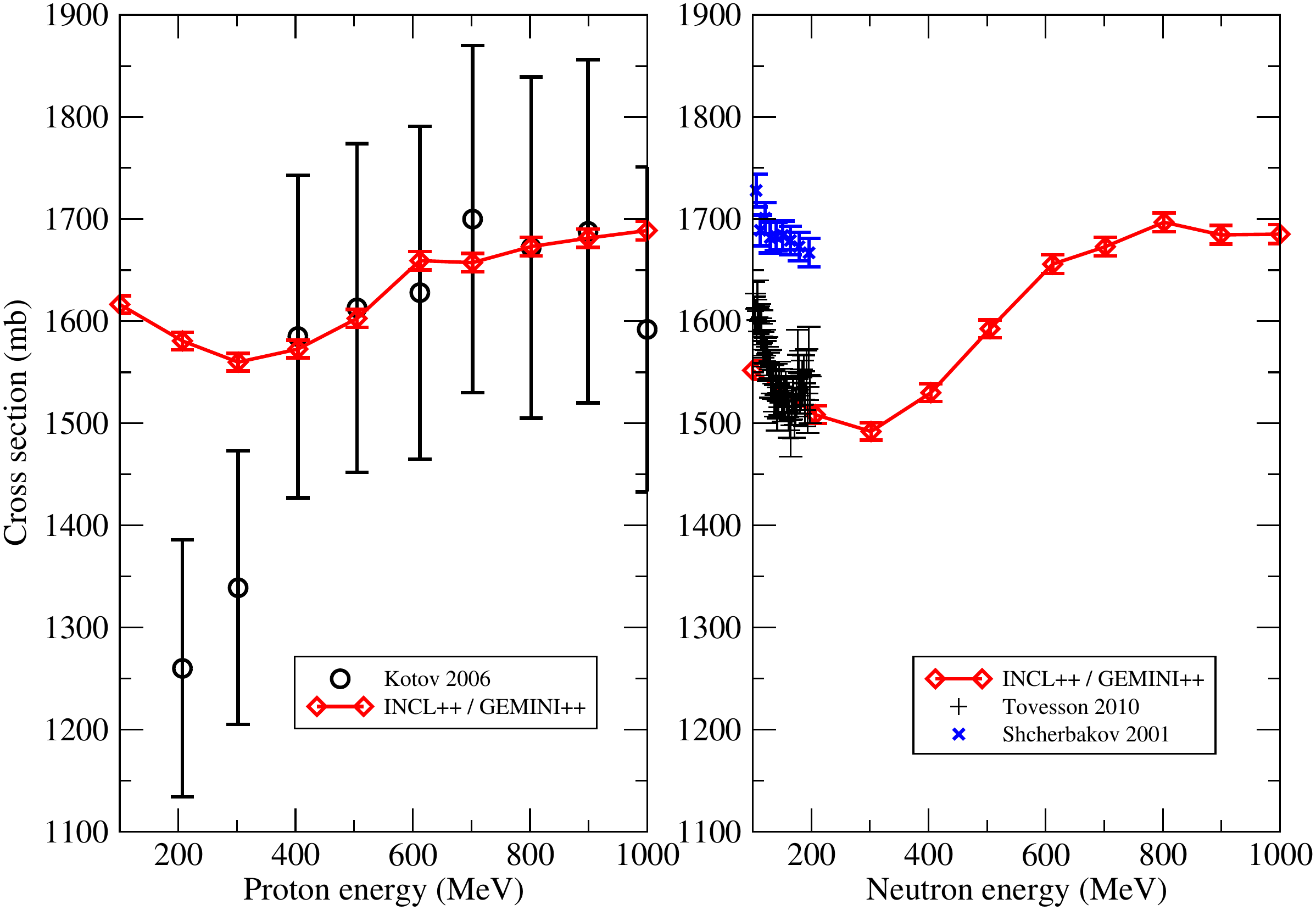}
	\caption{ $^{239}Pu$ fission cross sections. Experimental data are discussed in the text.}
	\label{label6}
\end{figure}

The fit of the $^{239}Pu(p,f)$ cross section, for which the only data in the energy range of interest are those of Kotov\cite{Ko06}, is characterized by a large $\chi_0^2$ value, because the two points at 207 MeV and 302 MeV are largely overestimated by our calculations, as shown in the left panel of Fig. \ref{label6}.

However, the same problem exists in the original fit of Ref.\cite{Ko06}, which clearly shows that a simultaneous reproduction of the two mentioned data and of those below 100 MeV from other experiments is not possible. The same difficulty 
had already been encountered for the $^{237}Np(p,f)$ reaction.

Using the same model parameters, \textit{i. e.} $\tilde a_f/\tilde a_n$ = 1.036 and $\Delta B_f$ = +0.3 MeV, in the $(n,f)$ calculations, shown in the right panel of Fig. \ref{label6}, yields a decent reproduction of the data of Ref.\cite{To10} (Tovesson 2010), while underestimating those of Ref.\cite{Sh01} (Shcherbakov 2001). It is to be stressed, however, that both measurements are relative to the $^{235}U(n,f)$ reaction and that the experimental ratios have been normalized with different $^{235}U(n,f)$ cross sections. In addition, the $(n,f)$ cross section of Ref.\cite{Sh01} appears to be larger than any reasonable extrapolation of the $(p,f)$ cross section down to the same energies, which is not expected on physical grounds. In any case, it is more meaningful to compare relative cross sections, as shown in the following sub-section.

\subsubsection{Lead - Bismuth}

In the lead-bismuth region, which is also very important for technological applications and is considered in the rest of this section, we have preferred to couple INCL++ with ABLA07, since the latter appears to produce more accurate cross section fits than GEMINI++. The fission model parameters we have
modified are the level density parameter at the saddle point, $a_f$, multiplied by a scale factor $k_f$, and the height of the liquid-drop fission barriers, shifted by a positive or negative quantity, $\Delta B_f$, as before. Table \ref{tabII} shows adopted parameters and $\chi^2$ tests for
proton induced fission of $^{208}Pb$, $^{nat}Pb$ and $^{209}Bi$.

\begin{table}[h]
	\begin{center}
		\begin{tabular}{|c|c|c|c|c|c|}\hline
			Isotope    & $Z^2/A$ & $k_f$ & $\Delta B_f$ (MeV) & $\chi^2_0$ & $Q\left(\chi^2_0\right)$ \\ \hline
			$^{208}Pb$ & 32.33   &   1.0 &      0.0           &   6.21     &       0.52               \\ \hline
			$^{nat}Pb$  & 32.45   &  0.995 &    +0.16          &  20.04     &       0.0055             \\ \hline
			$^{209}Bi$ & 32.96   &   1.01  &     +0.18           &   2.53      &       0.93                \\ \hline
		\end{tabular}
	\end{center}
	\caption{Model parameters and $\chi^2$ tests for proton-induced fission of lead and bismuth.}
	\label{tabII}
\end{table}  

The energy dependence of fission cross sections of lead and bismuth is quite different from that in the actinide region: at low energy, let us say 100 MeV, the $(n,f)$ cross section is 2 to 3 times smaller than the $(p,f)$ cross section and neither of them reaches a plateau at high energy, 1 GeV. Therefore, knowledge of the $(p,f)$ cross section in this energy range yields less useful information on the $(n,f)$ cross section with respect to actinides.

\subsubsection{$^{208}Pb$}

\begin{figure}[h]
	\centering
	\includegraphics[width=10cm]{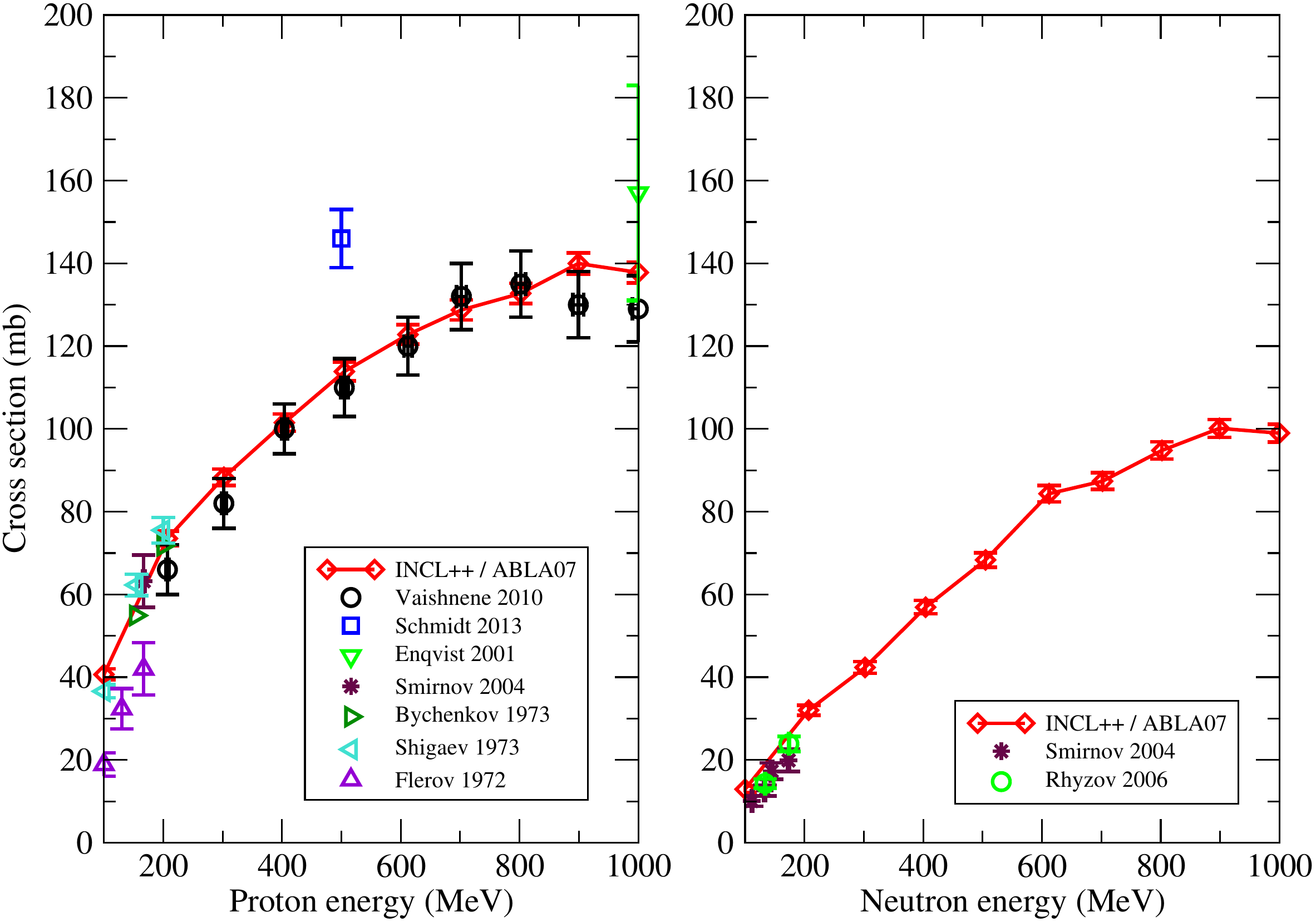}
	\caption{ $^{208}Pb$ fission cross sections. Experimental data are discussed in the text.}
	\label{label7}
\end{figure}

An experiment performed at the institute of Ref.\cite{Ko06} produced the $^{208}Pb(p,f)$ cross section at the same energies and the results, published in Ref.\cite{Va10} (Vaishnene 2010), are shown in the left panel of Fig.\ref{label7}, together with our calculations, which reproduce them satisfactorily with the default parameters of the
ABLA07 model, \textit{i. e.} $k_f$ = 1 and $\Delta B_f$ = 0. 

It is to be pointed out that $(p,f)$ data below 200 MeV are consistent with Ref.\cite{Va10}, with the
exception of Ref.\cite{Fl72} (Flerov 1972), whose data are somewhat lower than our calculation. At high energy, on the contrary, the recent result of Ref.\cite{Sc13} (Schmidt 2013) at 500 MeV as well as the value of Ref.\cite{En01} (Enqvist 2001) at 1 GeV are higher. Both experiments\cite{Sc13}\cite{En01} were performed in inverse kinematics. New data in this energy interval are needed in order to clarify the situation.


$(n,f)$ data are available up to 200 MeV from Refs.\cite{Sm04} ( Smirnov 2004) and \cite{Rh06} (Rhyzov 2006), which are in mutual agreement. They are reproduced in the right panel of Fig.\ref{label7} with the default value of $\tilde a_f$, at the cost of increasing the fission barriers by a quantity $\Delta B_f$ = 0.3 MeV.

\subsubsection{$^{nat}Pb$}

\begin{figure}[h]
	\centering
	\includegraphics[width=11cm]{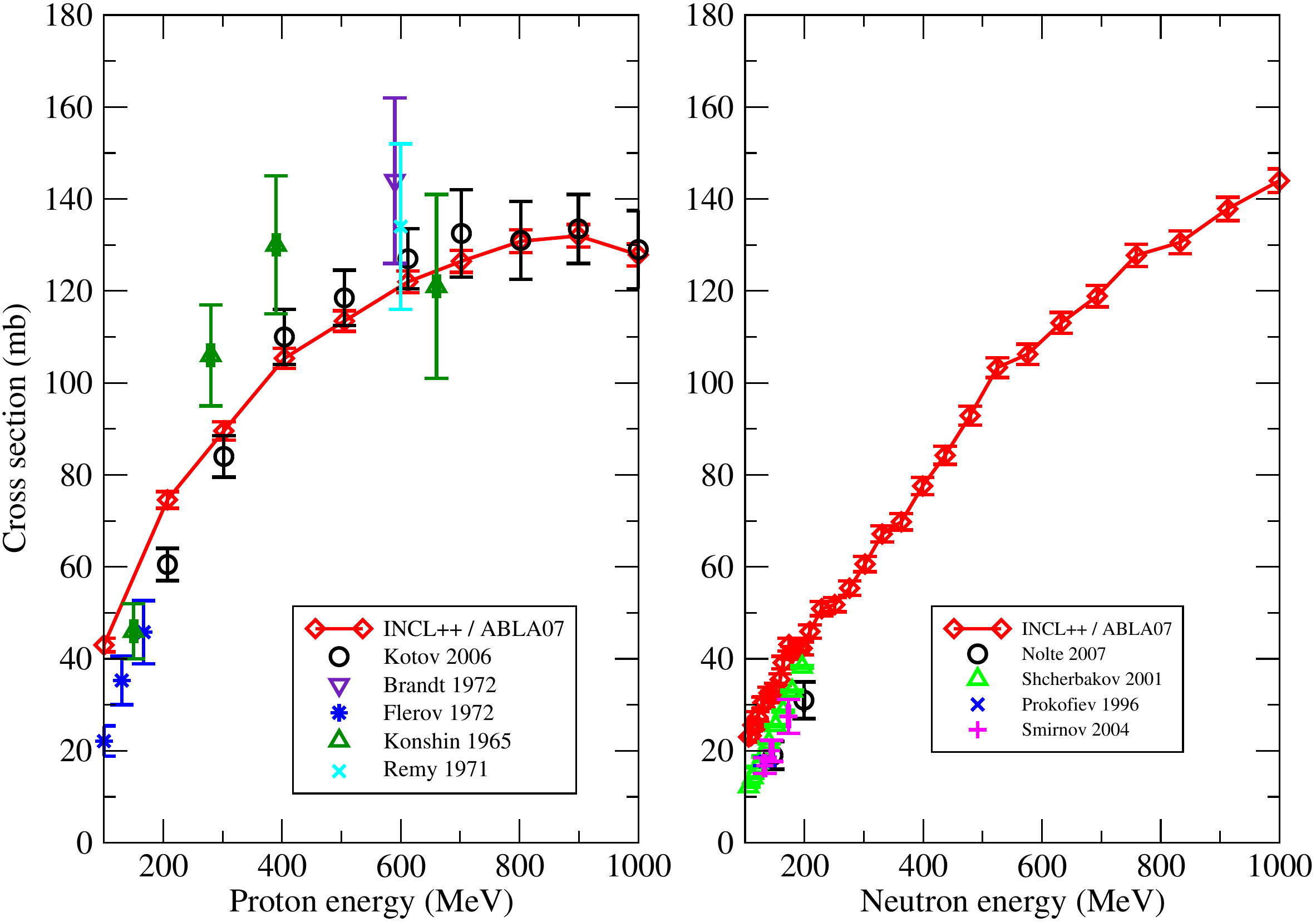}
	\caption{ $^{nat}Pb$ fission cross sections. Experimental data are discussed in the text.}
	\label{label8}
\end{figure}

Fission of natural lead is very important for applications to accelerator-driven systems, where the eutectic system acting as a spallation target and as a coolant consists of $^{nat}Pb$ and $^{209}Bi$.
As usual, Kotov\cite{Ko06} provides $(p,f)$ data from 200 MeV to 1 GeV. An older measurement (Flerov 1972)\cite{Fl72} below 200 MeV appears to be compatible with Kotov's data\cite{Ko06}, while other old experiments\cite{Ko65},\cite{Br72},\cite{Re71} (Koshin 1965, Brandt 1972 and Remy 1971, respectively) at higher energies seem to suggest a systematically higher cross section. As shown by the corresponding $\chi^2$ test in Table \ref{tabII}, our fit to the $(p,f)$ data of Kotov\cite{Ko06}, obtained by reducing the default $\tilde a_f$ parameter by a factor $k_f$ = 0.995 and by reducing the fission barrier by a quantity $\Delta B_f$ = -0.16 MeV is not very satisfactory; in particular, our calculations overestimate the low energy points, as shown in the left panel of Fig.\ref{label8}.

 The same difficulty appears also in $(n,f)$ calculations, carried out with $k_f$ = 1.005 and $\Delta B_f$ = 0, which overestimate the experimental data below 200 MeV, where there is good agreement of experiments (Prokofiev 1996)\cite{Pr96}, (Shcherbakov 2001)\cite{Sh01}, (Smirnov 2004)\cite{Sm04} , (Nolte 2007)\cite{No07}, as shown in the right panel of Fig. \ref{label8}. The relative $(n,f)$ data measured at n\_TOF up to 1 GeV\cite{Ta11} will be discussed in the next sub-section.


\subsubsection{$^{209}Bi$}

\begin{figure}[h]
	\centering
	\includegraphics[width=11cm]{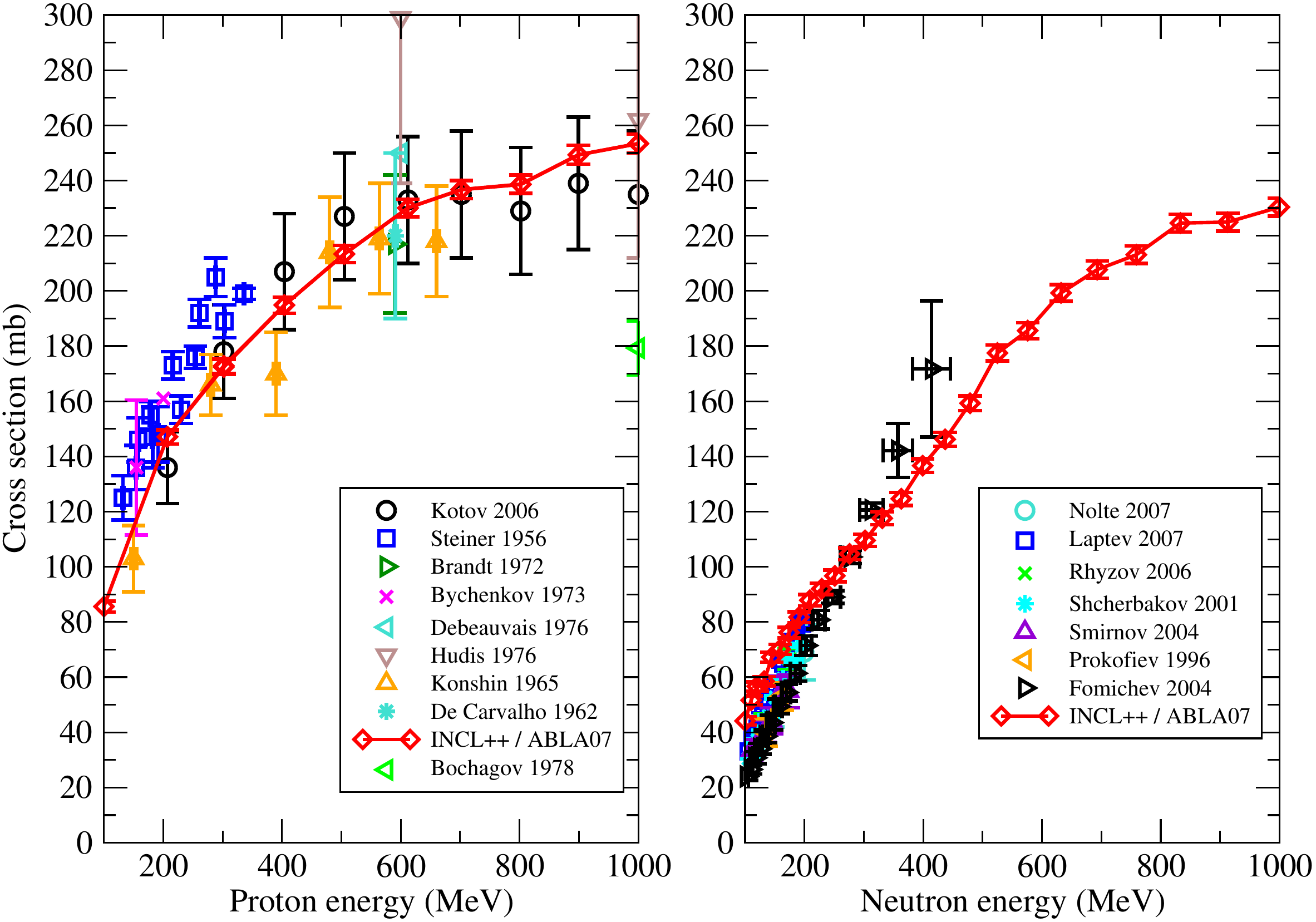}
	\caption{ $^{209}Bi$ fission cross sections. Experimental data are discussed in the text.}
	\label{label9}
\end{figure}

As far as the $(p,f)$ cross section is concerned, several measurements exist for $^{209}Bi$ and the majority of them (Steiner 1956)\cite{St56}, (De Carvalho 1962)\cite{DC62}, (Konshin 1965)\cite{Ko65}, (Brandt 1972)\cite{Br72}, (Bychenkov 1973)\cite{By73}, (Debeauvais 1976)\cite{DB76} are in reasonable agreement with our reference experiment Kotov 2006\cite{Ko06}, as shown in the left panel of Fig.\ref{label9}; the main exceptions are the point at 1 GeV from Ref.\cite{Bo78} (Bochagov 1978), which is much lower than the corresponding point from Ref.\cite{Ko06}, as well as  Ref.\cite{Hu76} (Hudis 1976), whose data are systematically higher than those of Ref.\cite{Ko06}.
 The latter are well reproduced in our calculations by multiplying the default $\tilde a_f$ parameter by a factor $k_f$ = 1.01 and increasing the fission barriers by an amount $\Delta B_f$ = +0.18 MeV.
Among the $(n,f)$ experiments, a key role is played by Refs.\cite{Pr96},\cite{No07} (Prokofiev 1996, Nolte 2007), which are relative to $(n,p)$ scattering and can be considered as measurements of the absolute $(n,f)$ cross section, while Refs\cite{Sh01} (Shcherbakov 2001), \cite{Fo04} (Fomichev 2004), \cite{La07} (Laptev 2007), \cite{Ta11} (Tarr\'io 2011) are relative to $^{235}U(n,f)$ and Refs.\cite{Sm04},\cite{Rh06} (Smirnov 2004 and Rhyzov 2006, respectively), are relative to $^{238}U$. Only Refs.\cite{Ta11},\cite{Fo04} extend above 200 MeV. Using in $(n,f)$ calculations the same parameters adopted in fitting $(p,f)$ data would underestimate the data
of Ref.\cite{Fo04} from 200 to 400 MeV, while reproducing the data of Ref.\cite{Ta11} relative to $^{235}U$ reasonably well. As a compromise between the two sets of data, we have adopted in $(n,f)$ calculations $k_f$ = 1.01 and $\Delta B_f$ = 0. More details will be given in the next sub-section.

\subsection{Relative Cross Sections}
This sub-section is dedicated to $(n,f)$ measurements relative to the $^{235}U(n,f)$ reaction, taken
as a standard, even if no reliable data of its absolute cross section exist above 200 MeV, with particular reference to n\_TOF\cite{Gu13} measurements up to 1 GeV. For those cases\cite{Pa10}\cite{Ta11} where the experimental ratios measured at n\_TOF have been normalized in the original publications to the $^{235}U(n,f)$ cross section from the JENDL/HE-2007 library a comparison of absolute cross sections has been added to that of relative cross sections. The parameters adopted in the calculations are those given for the $(n,f)$ reactions in the sub-section on absolute cross sections.

The cross section ratio to be compared with measurements is the ratio of theoretical cross sections obtained with our Monte Carlo simulations, which make it possible to associate to any $\sigma(E)$ an
uncertainty $\Delta \sigma(E)$ attributable to counting statistics. If $R = \sigma_1 / \sigma_2$ is the ratio in question, the relative error $\Delta R/R$ is obtained by the quadratic law of propagation of relative errors of two uncorrelated variables $\sigma_1$ and $\sigma_2$
\begin{equation}
\frac{\Delta R}{R} =\left[\left(\frac{\Delta \sigma_1}{\sigma_1}\right)^2 + \left(\frac{\Delta \sigma_2}{\sigma_2} \right)^2\right]^{1/2} \quad .
\label{ratio}
\end{equation}

\subsubsection{$^{238}U$}

\begin{figure}[!h]
	\centering
	\includegraphics[width=11cm]{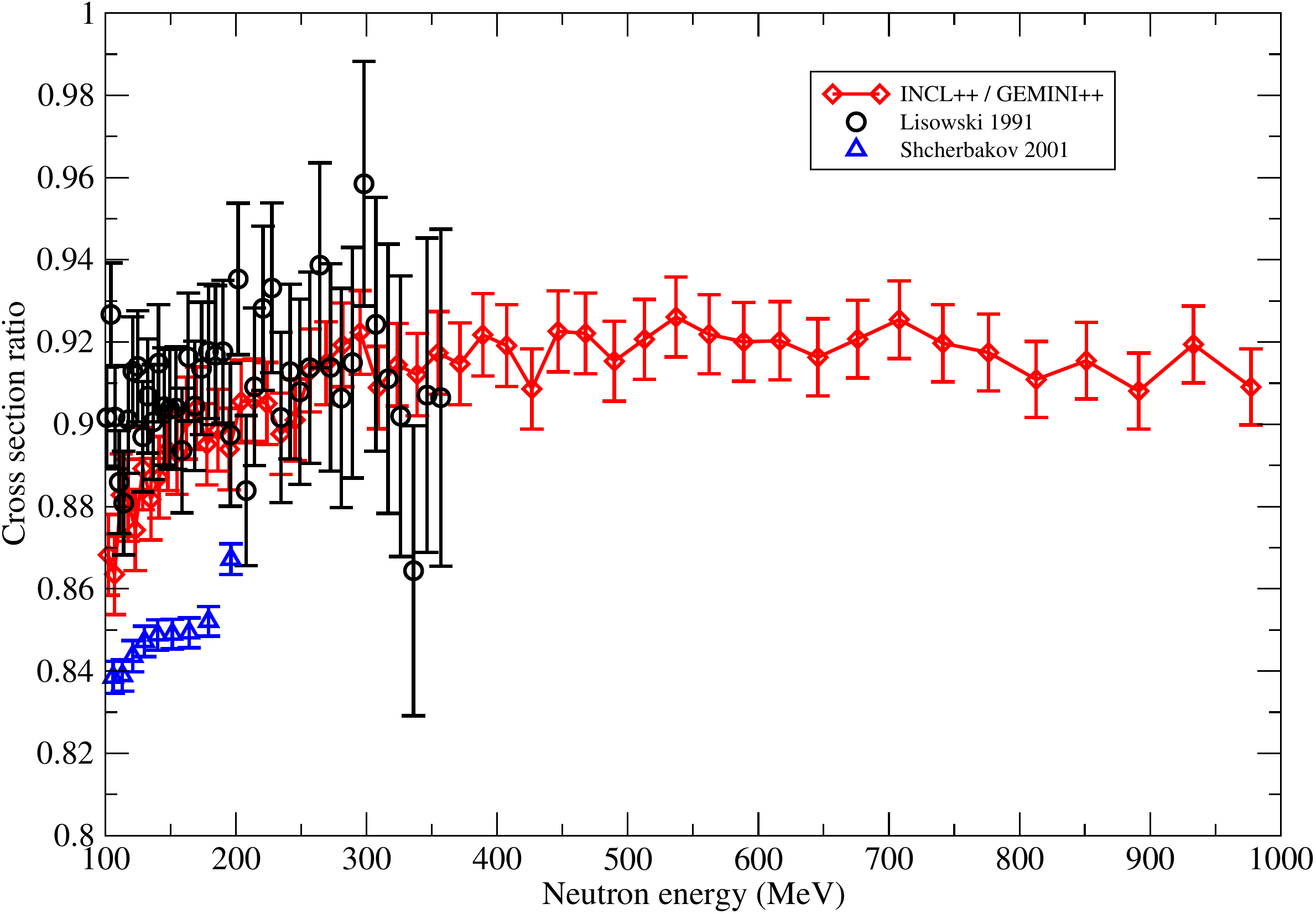}
	\caption{ Ratio of $^{238}U(n,f)$ to $^{235}U(n,f)$ fission cross section. Experimental data are discussed in the text.}
	\label{label10}
\end{figure}

Figure \ref{label10} compares calculated and experimental ratios for the $^{238}U(n,f)$ reaction.

Our results agree with Ref.\cite{Li91}, which still represents one of the most reliable measurements
of $(n,f)$ cross sections in the intermediate energy region, and are somewhat higher than Ref.\cite{Sh01}. Preliminary n\_TOF results up to 1 GeV, not reported in Fig. \ref{label10}, compare reasonably
well with our calculations.

\subsubsection{$^{234}U$}

\begin{figure}[h]
	\centering
	\includegraphics[width=10cm]{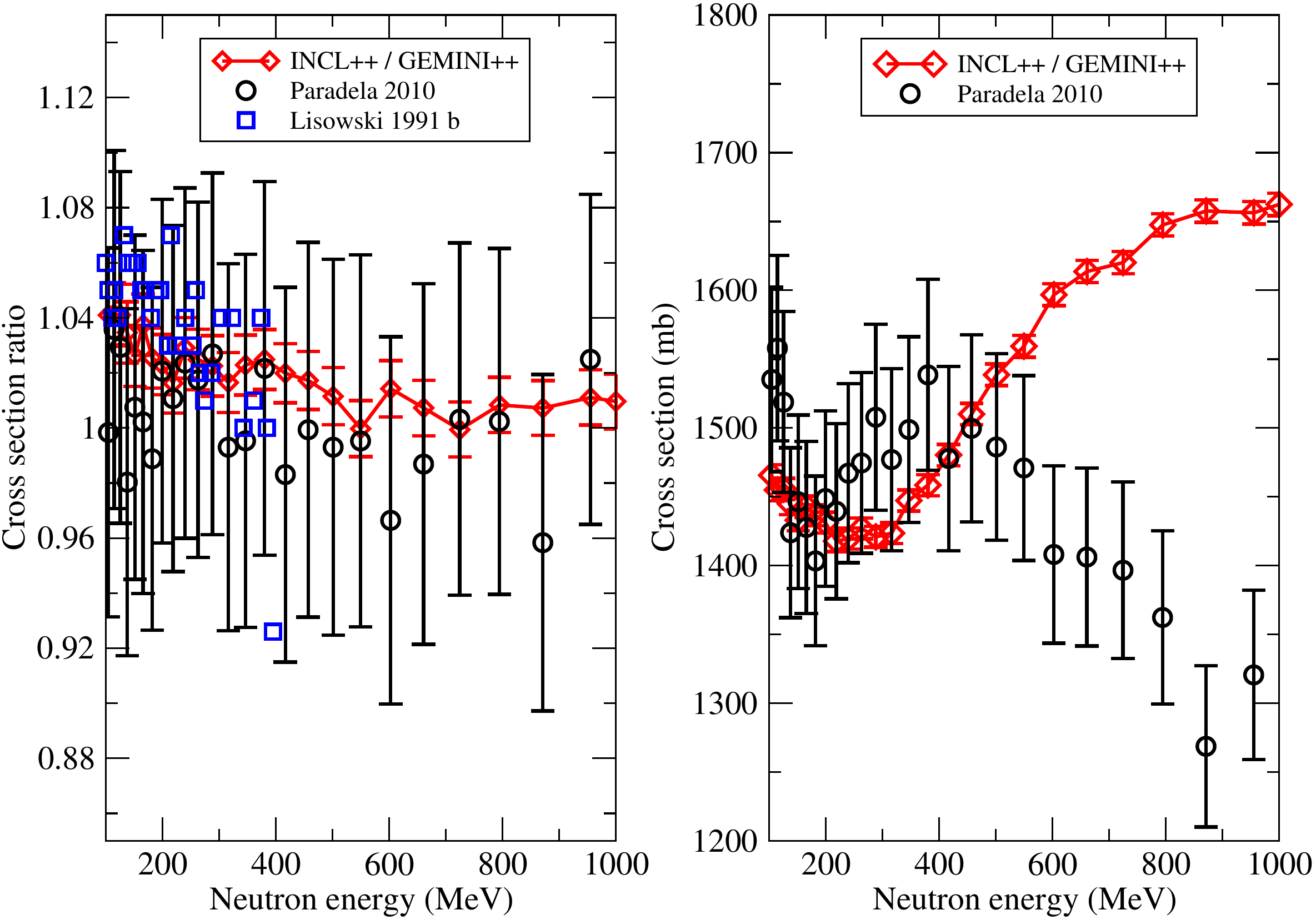}
	\caption{Left panel: $(n,f)$ cross section of $^{234}U$ relative to $^{235}U$. Right panel: absolute $(n,f)$ cross section of $^{234}U$. Experimental data are discussed in the text.}
	\label{label11}
\end{figure}

No $(p,f)$ data are available for $^{234}U$, whose $(n,f)$ cross section relative to $^{235}U$ has been measured up to 400 MeV in Ref.\cite{Li91b} (Lisowski 1991 b) and at the n\_TOF facility up to 1 GeV using parallel plate avalanche counters in Ref.\cite{Pa10} (Paradela 2010) and up to 200 MeV using fission ionization chambers in Ref.\cite{Ka14}; the two n\_TOF experiments are in good mutual agreement, therefore we take into account only the more extended measurement\cite{Pa10} for comparison with our calculations, done with the
same fission parameters as $^{235}U$, namely $\tilde a_f/\tilde a_n$ = 1.050 and $\Delta B_f$ - 0.2 MeV.
The left panel of Fig.\ref{label11} compares theoretical and experimental ratios, in overall good agreement, while the right panel compares our absolute cross section with the one obtained in Ref.\cite{Pa10} by normalizing the experimental ratio to the JENDL/HE-2007 evaluation of the $^{235}U(n,f)$ cross section.
As one can see from the right panel of Fig. \ref{label1} the disagreement increases with increasing neutron energy, because our calculated $^{235}U(n,f)$ cross section shows a large plateau at high energy, while the JENDL evaluation steadily decreases in the same energy range.

\subsubsection{$^{233}U$}

\begin{figure}[h]
	\centering
	\includegraphics[width=12cm]{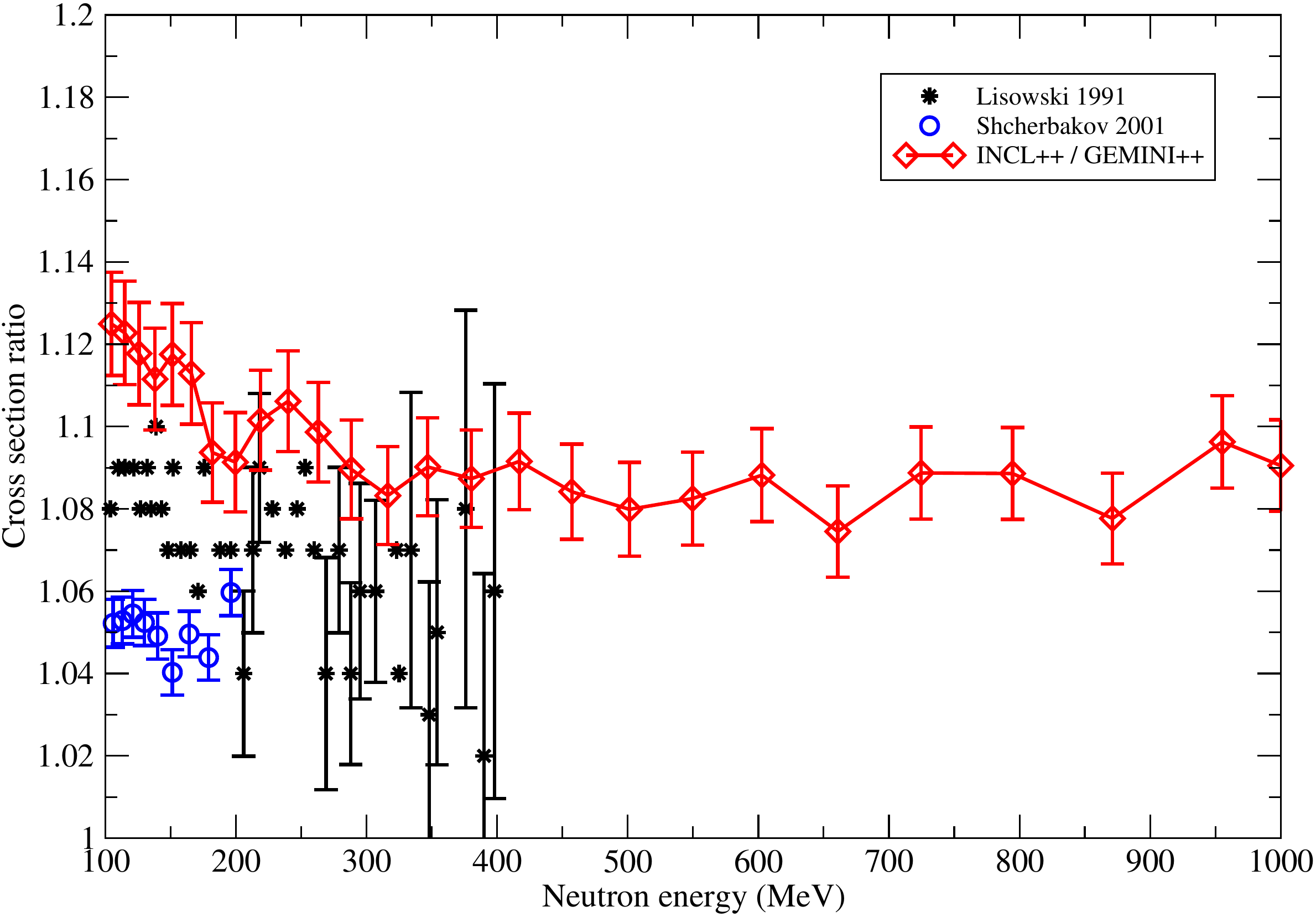}
	\caption{ $(n,f)$ cross section of $^{233}U$ relative to $^{235}U$. Experimental data are discussed in the text.}
	\label{label12}
\end{figure}

A  somewhat larger value of $\tilde a_f/\tilde a_n$ is necessary for reproduction of the fission measurements of $^{233}U$. Fig.\ref{label12} shows a comparison of available 
data relative to $^{235}U$ from Refs.\cite{Li91b},\cite{Sh01} with the theoretical cross section ratio obtained with $\tilde a_f/\tilde a_n$ = 1.10 and $\Delta B_f$ = 0. A positive value of $\Delta B_f$ would further bring the relative cross section to better agreement with experimental data. New data up to 1 GeV are expected in the near future from the n\_TOF experiment.

\subsubsection{$^{232}Th$}

\begin{figure}[!h]
	\centering
	\includegraphics[width=12cm]{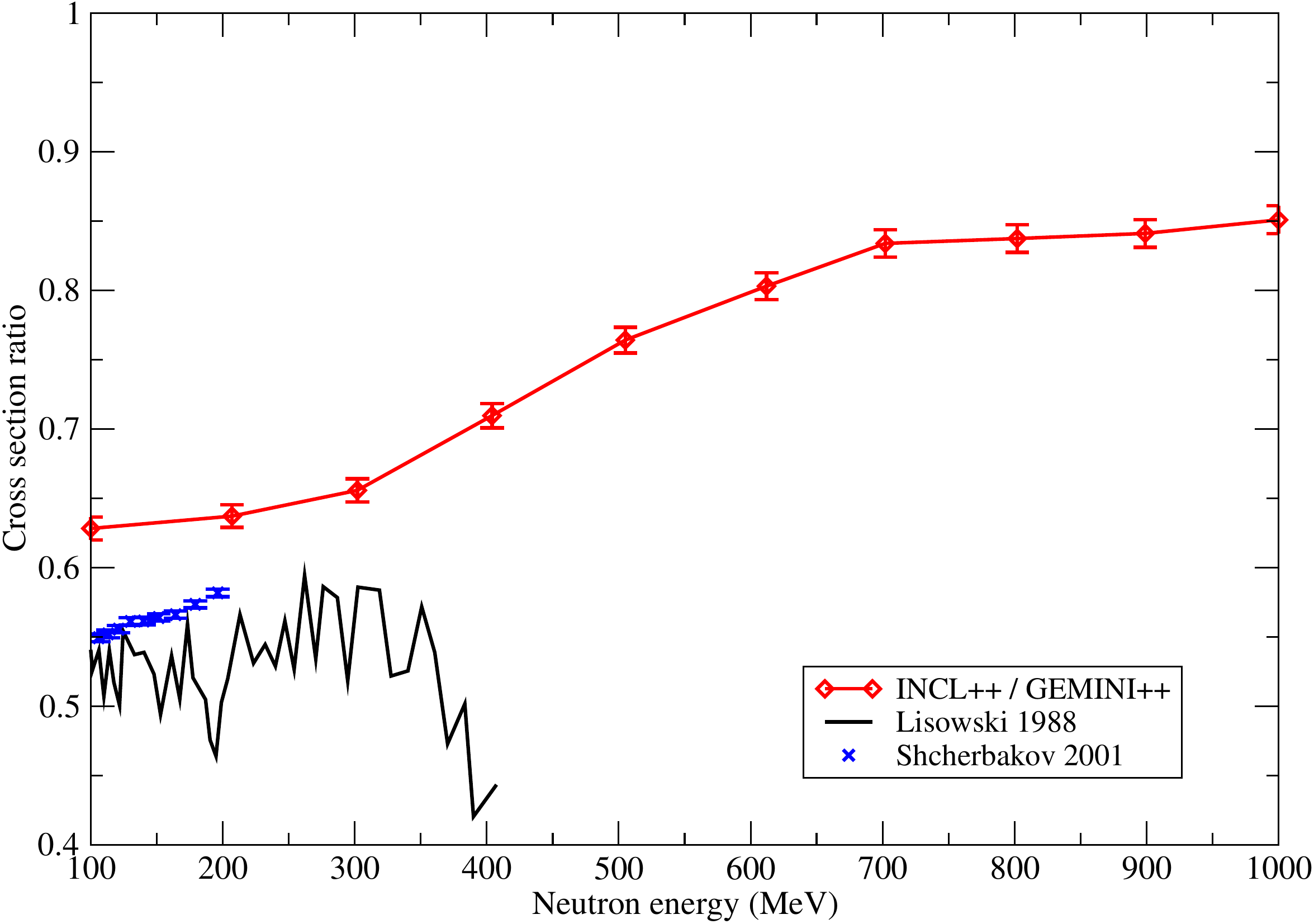}
	\caption{ $(n,f)$ cross section of $^{232}Th$ relative to $^{235}U$. Experimental data are discussed in the text.}
	\label{label13}
\end{figure}

As already discussed in the preceding sub-section, the n\_TOF results for the $(n,f)$ cross section of $^{232}Th$ are still preliminary and a more reliable $^{235}U(n,f)$ cross section in the intermediate energy range is recommended for their normalization. The difficulty with the present data\cite{Pa06} clearly appears in the comparison of our calculated cross-section ratio, which reproduces the $(n,f)$ values of Fig \ref{label4}, with the presently available experimental ratios\cite{Li88}\cite{Sh01}, which are in reasonable mutual agreement and significantly lower than our results, as shown in Fig.\ref{label13}. 

\subsubsection{$^{237}Np$}

\begin{figure}[!h]
	\centering
	\includegraphics[width=9cm]{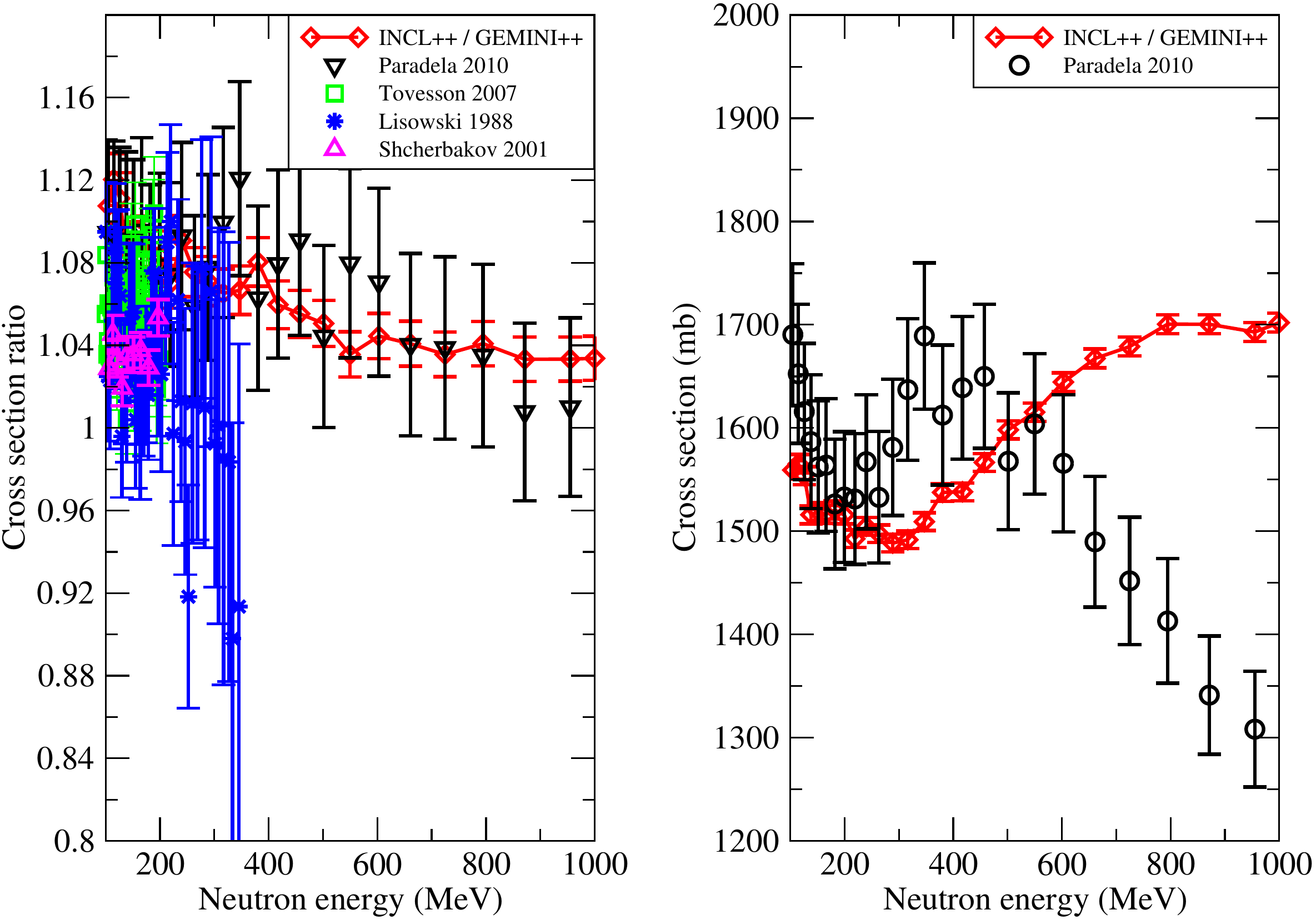}
	\caption{ Left panel: $(n,f)$ cross section of $^{237}Np$ relative to $^{235}U$. Right panel: absolute $(n,f)$ cross section of $^{237}Np$. Experimental data are discussed in the text.}
	\label{label14}
\end{figure}

The left panel of Fig.\ref{label14} compares the ratios of the $^{237}Np(n,f)$ cross section to the $^{235}U(n,f)$ cross section measured at n\_TOF\cite{Pa10} with our calculated results, which turn out to be in overall agreement with the experiment.  
On the contrary, the right panel shows a significant disagreement of the corresponding absolute cross sections in the high energy region, owing to the normalization to the JENDL/HE-2007 evaluation of $^{235}U(n,f)$ adopted in Ref.\cite{Pa10}.

\subsubsection{$^{239}Pu$}

\begin{figure}[!h]
	\centering
	\includegraphics[width=9cm]{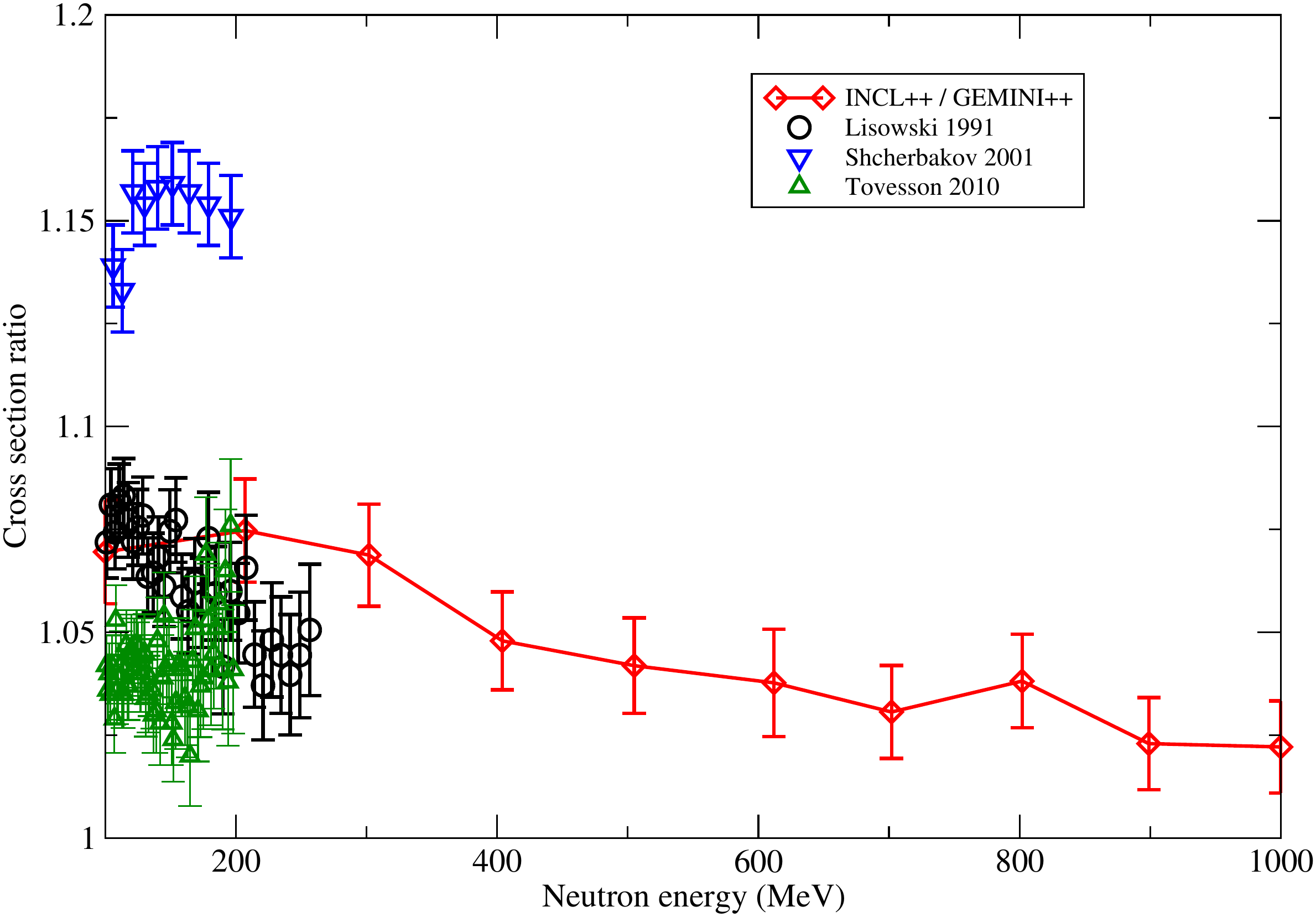}
	\caption{ $(n,f)$ cross section of $^{239}Pu$ relative to $^{235}U$. Experimental data are discussed in the text.}
	\label{label15}
\end{figure}

The relative $(n,f)$ cross section of $^{239}Pu$, shown in Fig. \ref{label15}, compares reasonably well with the data of Ref.\cite{Li91} below 150 MeV and with those of Ref.\cite{To10} between 150 and 200 MeV. 
The two series of data partially overlap, while those of Ref.\cite{Sh01} are systematically higher than them and than our calculations as well.

\subsubsection{$^{nat}Pb$}

\begin{figure}[h]
	\centering
	\includegraphics[width=10cm]{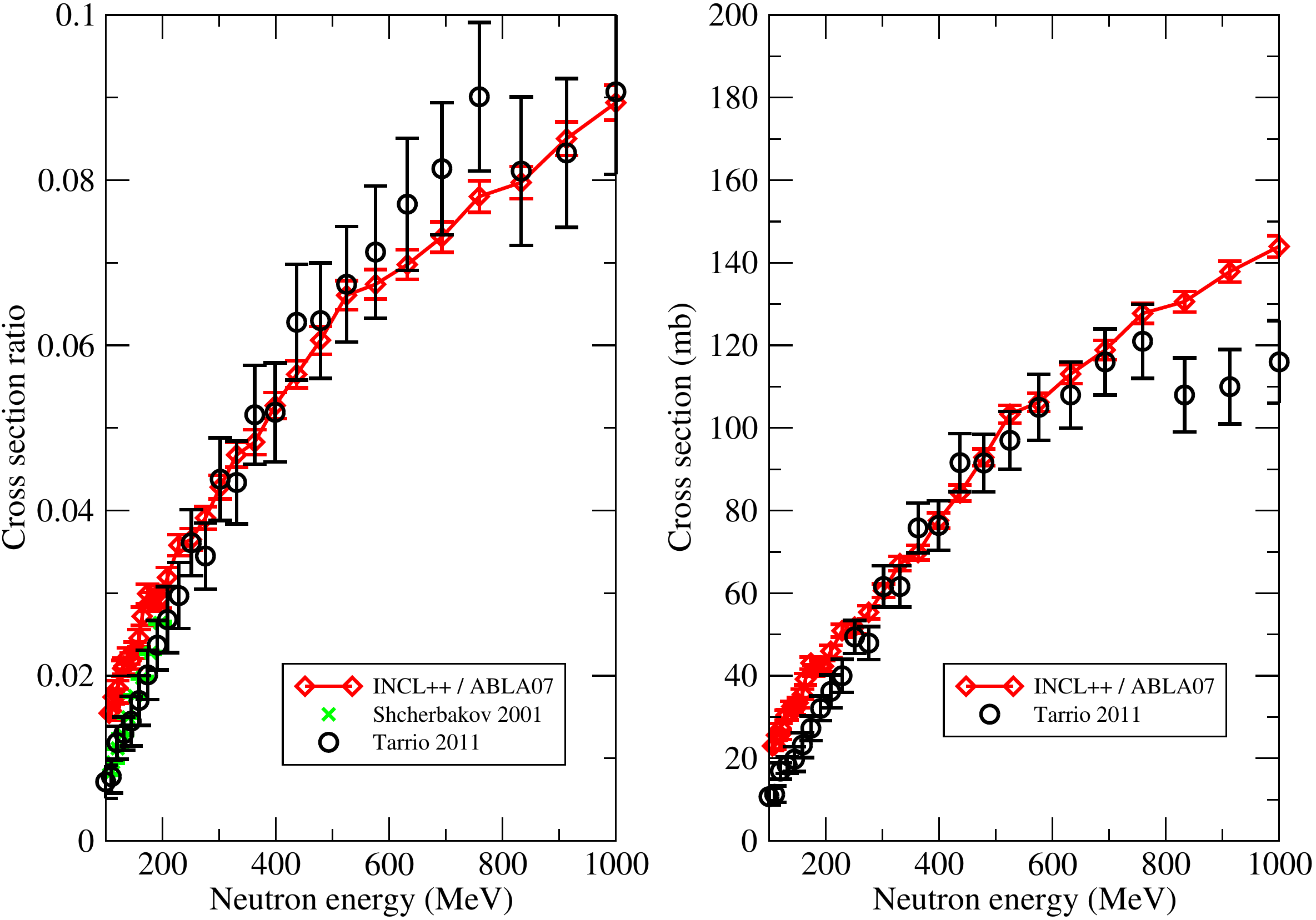}
	\caption{Left panel: $(n,f)$ cross section of $^{nat}Pb$ relative to $^{235}U$. Right panel: absolute $(n,f)$ cross section of $^{nat}Pb$. Experimental data are discussed in the text.}
	\label{label16}
\end{figure}

The $(n,f)$ cross section of $^{nat}Pb$ relative to $^{235}U$ has been measured at n\_TOF up to 1 GeV\cite{Ta11}.
The left panel of Fig.\ref{label16} shows a comparison of experimental and theoretical cross section ratios, computed by increasing the default $a_f$ parameter by a factor of 1.005, without any change of fission barriers. As already noticed for the absolute $(n,f)$ cross section discussed in the preceding sub-section, the calculations overestimate the experimental data at low energies. For the sake of completeness, the right panel of Fig. \ref{label16} compares the calculated absolute cross section with the one obtained by normalizing the relative cross section measured at n\_TOF to the evaluated $^{235}U(n,f)$ cross section of the JENDL/HE-2007 library. The disagreement at high energies is mainly due to the difference of the two computed $^{235}U$ cross sections.

\subsubsection{$^{209}Bi$}

\begin{figure}[!h]
	\centering
	\includegraphics[width=12cm]{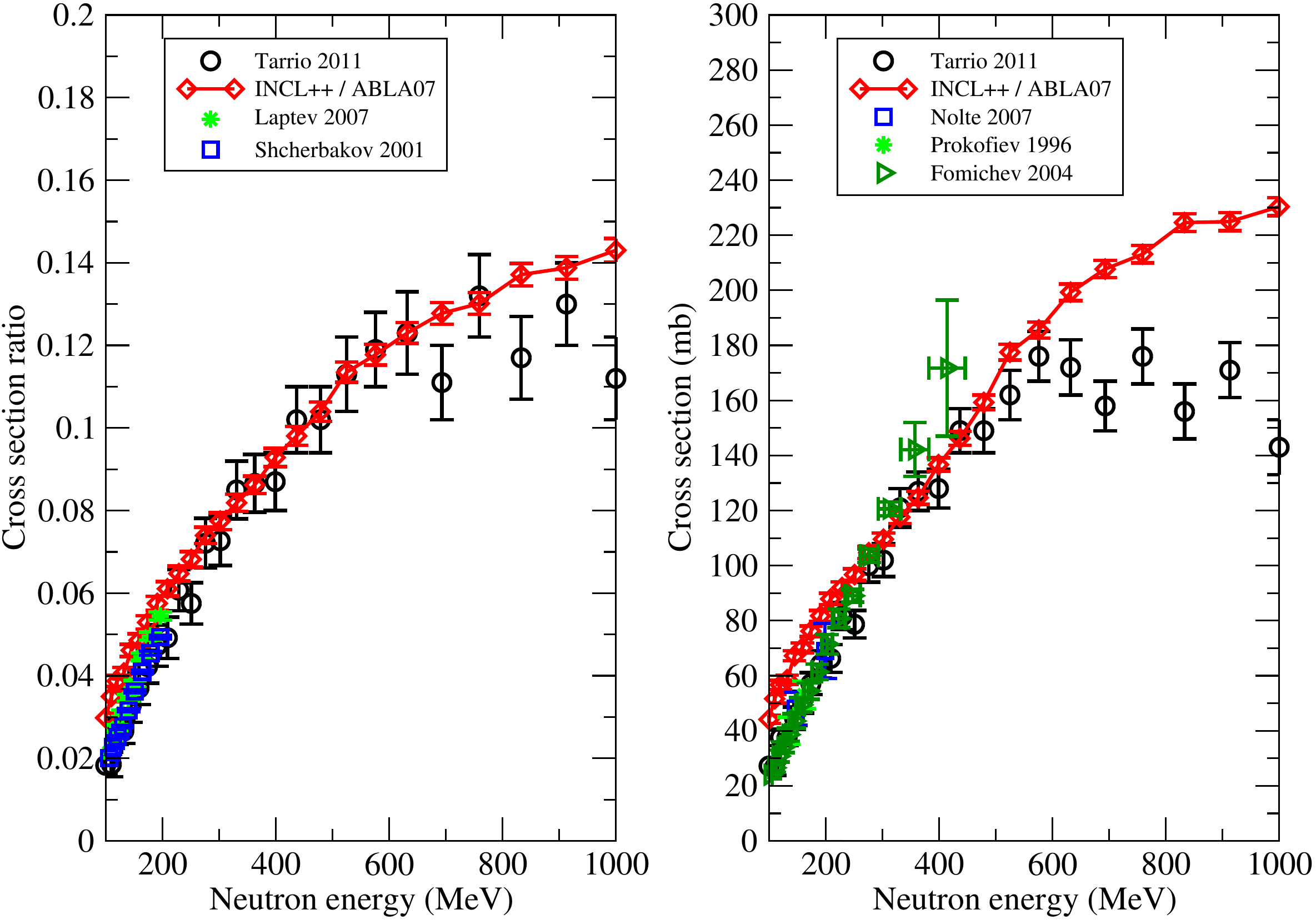}
	\caption{Left panel: $(n,f)$ cross section of $^{209}Bi$ relative to $^{235}U$. Right panel: absolute $(n,f)$ cross section of $^{209}Bi$. Experimental data are discussed in the text.}
	\label{label17}
\end{figure}

Finally, Fig. \ref{label17} shows the results of our $(n,f)$ calculations for $^{209}Bi$, which slightly overestimate the data of Ref.\cite{Ta11} relative to $^{235}U$ below 200 MeV, as shown in the left panel of the figure. 

This is the result of a compromise that takes
into account the absolute data of Ref.\cite{Fo04} up to 400 MeV, shown in the right panel of the figure together with the relative data of Ref.\cite{Ta11} normalized to the JENDL/HE-2007 cross section of the $^{235}U(n,f)$ reaction. Also the measurement of Ref.\cite{Fo04} is relative to $^{235}U$, but we do not know the values adopted for normalization, so that we could consider only the absolute cross section. The disagreement with Ref.\cite{Ta11} above 500 MeV is mainly due to the JENDL/HE-2007 cross section adopted in that reference for normalization.

Summing up, $(p,f)$ and $(n,f)$ cross sections have been reproduced with the same model parameters in the cases of $^{235,238}U$, $^{237}Np$, $^{239}Pu$ and $^{208}Pb$; the same level density parameters, but different corrections to the heights of fission barriers have been adopted for $^{233}U$, $^{232}Th$ and $^{209}Bi$; finally, slightly different level density parameters and barrier height corrections have been used for $^{nat}Pb$: this is a peculiar case, since the natural element consists of sizable abundances of $^{208}Pb$, $^{207}Pb$ and $^{206}Pb$. In principle, each isotope should have its own model parameters, but, for the sake of simplicity, we have chosen to treat $^{nat}Pb$ as an effective nucleus.

\section{Conclusions and Outlook}
In the present work we have reproduced $(p,f)$ cross sections in the intermediate energy range for a number of actinides and pre-actinides relevant to nuclear energy applications by a two-parameter fit:
an increase of the level density parameter at the saddle point with increasing fissility parameter is
observed in specific isotopic chains, such as uranium nuclei in Table I and lead nuclei in Table II, but the present data are not sufficient to determine a global trend in either region. Therefore, the results of the present work cannot be used for prediction of fission cross sections of isotopes for which neither $(p,f)$ or $(n,f)$ data exist in the intermediate energy range.

The model parameters that reproduce $(p,f)$ cross sections have been used as a first guess in the calculation of $(n,f)$ cross sections, where the comparison with experiments is complicated by the fact that the large majority of data are relative to fission standards, such as $^{235,238}U$ in the actinide
region and/or $^{209}Bi$ in the pre-actinide region, but even for the standards absolute cross section data are not available above 200 MeV. $(n,f)$ cross sections are systematically lower than $(p,f)$ cross sections around 100 MeV, but the differences are expected to reduce with increasing projectile energy, particularly for actinides, whose $(p,f)$ data exhibit a large plateau from 500 MeV to 1 GeV. This effect is clearly seen in our calculations.

As far as $(n,f)$ cross sections are concerned, the only relative measurements extending up to 1 GeV, due to the n\_TOF collaboration, are satisactorily reproduced by our calculations. The disagreement between corresponding absolute cross sections is almost entirely due to the normalization to the JENDL/HE-2007 evaluation of the $^{235}U(n,f)$ reaction adopted in n\_TOF papers\cite{Pa10},\cite{Ta11}.

A deeper comparison with evaluated data libraries such as JENDL/HE-2007 would require an extension of the energy range of our calculations: as already shown in the case of $^{nat}U$, whose calculations up to 3 GeV look promising, the codes used in this work allow us to extend cross section calculations up to 3 GeV, but moving the lower extremum down to 20 MeV could only be done by resorting to fully quantum-mechanical models, since the very concept of intranuclear cascade model loses its applicability at low energies.

A recent example of consistent calculations of $(p,f)$ and $(n,f)$ calculations in the energy range from 20 MeV to 1 GeV for a number of actinides is provided by Ref.\cite{Gr13}, where use is made of the MCFx system of codes, permitting a good reproduction of many experimental data at the cost of adjusting the heights of the fission barriers of remnants obtained from microscopic calculations. This has been done by us, too,  at the very phenomenologic level permitted by the $\Delta B_f$ parameter of Tables I-II.

An extension of our fission calculations to the energy range from 20 MeV to 3 GeV is planned for the nuclei already suggested as fission standards, $^{235,238}U$ and $^{209}Bi$, with the specific scope of support to the analysis of $(n,f)$ cross sections measured at the n\_TOF facility\cite{Gu13} at CERN.

\section*{Acknowledgements}
We acknowledge valuable discussions with Drs. Sylvie Leray, Aleksandra Keli\'{c}-Heil, Nicola Colonna, Carlos Paradela and Laurent Tassan-Got.



\end{document}